\documentclass[ALICE,manyauthors]{cernphprep}
\usepackage[comma,square,numbers,sort&compress]{natbib}
\usepackage{hyperref}
\usepackage{lineno}
\usepackage{xspace}
\usepackage{color}
\usepackage{here}
\usepackage[utf8]{inputenc}
\DeclareUnicodeCharacter{2010}{-}
\usepackage{orcidlink}
\begin{document}
%

\newcommand{\pp}           {pp\xspace}
\newcommand{\ppbar}        {\mbox{$\mathrm {p\overline{p}}$}\xspace}
\newcommand{\XeXe}         {\mbox{Xe--Xe}\xspace}
\newcommand{\PbPb}         {\mbox{Pb--Pb}\xspace}
\newcommand{\pA}           {\mbox{pA}\xspace}
\newcommand{\pPb}          {\mbox{p--Pb}\xspace}
\newcommand{\AuAu}         {\mbox{Au--Au}\xspace}
\newcommand{\dAu}          {\mbox{d--Au}\xspace}
\newcommand{\pAu}          {\mbox{p--Au}\xspace}
\newcommand{\dAl}          {\mbox{p--Al}\xspace}
\newcommand{\HeAu}         {\mbox{He--Au}\xspace}

\newcommand{\s}            {\ensuremath{\sqrt{s}}\xspace}
\newcommand{\snn}          {\ensuremath{\sqrt{s_{\mathrm{NN}}}}\xspace}
\newcommand{\pt}           {\ensuremath{p_{\rm T}}\xspace}
\newcommand{\meanpt}       {$\langle p_{\mathrm{T}}\rangle$\xspace}
\newcommand{\ycms}         {\ensuremath{y_{\rm CMS}}\xspace}
\newcommand{\ylab}         {\ensuremath{y_{\rm lab}}\xspace}
\newcommand{\etarange}[1]  {\mbox{$\left | \eta \right |~<~#1$}}
\newcommand{\yrange}[1]    {\mbox{$\left | y \right |~<~#1$}}
\newcommand{\dndy}         {\ensuremath{\mathrm{d}N_\mathrm{ch}/\mathrm{d}y}\xspace}
\newcommand{\dndeta}       {\ensuremath{\mathrm{d}\it{N}_\mathrm{ch}/\mathrm{d}\eta}\xspace}
\newcommand{\avdndeta}     {\ensuremath{\langle\dndeta\rangle}\xspace}
\newcommand{\dNdy}         {\ensuremath{\mathrm{d}N_\mathrm{ch}/\mathrm{d}y}\xspace}
\newcommand{\Npart}        {\ensuremath{N_\mathrm{part}}\xspace}
\newcommand{\Ncoll}        {\ensuremath{N_\mathrm{coll}}\xspace}
\newcommand{\dEdx}         {\ensuremath{\textrm{d}E/\textrm{d}x}\xspace}
\newcommand{\RpPb}         {\ensuremath{R_{\rm pPb}}\xspace}

\newcommand{\nineH}        {$\sqrt{s}~=~0.9$~Te\kern-.1emV\xspace}
\newcommand{\seven}        {$\sqrt{s}~=~7$~Te\kern-.1emV\xspace}
\newcommand{\twoH}         {$\sqrt{s}~=~0.2$~Te\kern-.1emV\xspace}
\newcommand{\twosevensix}  {$\sqrt{s}~=~2.76$~Te\kern-.1emV\xspace}
\newcommand{\five}         {$\sqrt{s}~=~5.02$~Te\kern-.1emV\xspace}
\newcommand{\twosevensixnn}{$\sqrt{s_{\mathrm{NN}}}~=~2.76$~Te\kern-.1emV\xspace}
\newcommand{\fivenn}       {$\sqrt{s_{\mathrm{NN}}}~=~5.02$~Te\kern-.1emV\xspace}
\newcommand{\LT}           {L{\'e}vy-Tsallis\xspace}
\newcommand{\GeVc}         {Ge\kern-.1emV/$c$\xspace}
\newcommand{\MeVc}         {Me\kern-.1emV/$c$\xspace}
\newcommand{\TeV}          {Te\kern-.1emV\xspace}
\newcommand{\GeV}          {Ge\kern-.1emV\xspace}
\newcommand{\MeV}          {Me\kern-.1emV\xspace}
\newcommand{\GeVmass}      {Ge\kern-.2emV/$c^2$\xspace}
\newcommand{\MeVmass}      {Me\kern-.2emV/$c^2$\xspace}
\newcommand{\lumi}         {\ensuremath{\mathcal{L}}\xspace}

\newcommand{\ITS}          {\rm{ITS}\xspace}
\newcommand{\TOF}          {\rm{TOF}\xspace}
\newcommand{\ZDC}          {\rm{ZDC}\xspace}
\newcommand{\ZDCs}         {\rm{ZDCs}\xspace}
\newcommand{\ZNA}          {\rm{ZNA}\xspace}
\newcommand{\ZNC}          {\rm{ZNC}\xspace}
\newcommand{\SPD}          {\rm{SPD}\xspace}
\newcommand{\SDD}          {\rm{SDD}\xspace}
\newcommand{\SSD}          {\rm{SSD}\xspace}
\newcommand{\TPC}          {\rm{TPC}\xspace}
\newcommand{\TRD}          {\rm{TRD}\xspace}
\newcommand{\VZERO}        {\rm{V0}\xspace}
\newcommand{\VZEROA}       {\rm{V0A}\xspace}
\newcommand{\VZEROC}       {\rm{V0C}\xspace}
\newcommand{\Vdecay} 	   {\ensuremath{V^{0}}\xspace}

\newcommand{\ee}           {\ensuremath{e^{+}e^{-}}} 
\newcommand{\pip}          {\ensuremath{\pi^{+}}\xspace}
\newcommand{\pim}          {\ensuremath{\pi^{-}}\xspace}
\newcommand{\kap}          {\ensuremath{\rm{K}^{+}}\xspace}
\newcommand{\kam}          {\ensuremath{\rm{K}^{-}}\xspace}
\newcommand{\pbar}         {\ensuremath{\rm\overline{p}}\xspace}
\newcommand{\kzero}        {\ensuremath{{\rm K}^{0}_{\rm{S}}}\xspace}
\newcommand{\lmb}          {\ensuremath{\Lambda}\xspace}
\newcommand{\almb}         {\ensuremath{\overline{\Lambda}}\xspace}
\newcommand{\Om}           {\ensuremath{\Omega^-}\xspace}
\newcommand{\Mo}           {\ensuremath{\overline{\Omega}^+}\xspace}
\newcommand{\X}            {\ensuremath{\Xi^-}\xspace}
\newcommand{\Ix}           {\ensuremath{\overline{\Xi}^+}\xspace}
\newcommand{\Xis}          {\ensuremath{\Xi^{\pm}}\xspace}
\newcommand{\Oms}          {\ensuremath{\Omega^{\pm}}\xspace}
\newcommand{\degree}       {\ensuremath{^{\rm o}}\xspace}

\newcommand{\Veta}        {\ensuremath{v_{2}(\eta)}\xspace}
\newcommand{\vtwo}          {\ensuremath{v_{2}}\xspace}
\newcommand{\FMD}          {\rm{FMD}\xspace}
\newcommand{\FMDa}          {\rm{FMD1}\xspace}
\newcommand{\FMDab}          {\rm{FMD1,2}\xspace}
\newcommand{\FMDb}          {\rm{FMD2}\xspace}
\newcommand{\FMDc}          {\rm{FMD3}\xspace}
\newcommand{\DCA}          {\rm{DCA}\xspace}

\begin{titlepage}
\PHyear{2023}       
\PHnumber{196}      
\PHdate{30 August}  

\title{Measurements of long-range two-particle correlation over a wide pseudorapidity range in \pPb collisions at \texorpdfstring{$\sqrt{s_{\mathbf{NN}}}$~=~5.02~\TeV}{}}
\ShortTitle{Long-range two-particle correlation over a wide $\eta$ range in \pPb collisions}   

\Collaboration{ALICE Collaboration\thanks{See Appendix~\ref{app:collab} for the list of collaboration members}}
\ShortAuthor{ALICE Collaboration} 

\begin{abstract}
Correlations in azimuthal angle extending over a long range in pseudorapidity between particles, usually called the ``ridge'' phenomenon, were discovered in heavy-ion collisions, and later found in \pp and \pPb collisions. In large systems, they are thought to arise from the expansion (collective flow) of the produced particles. 
Extending these measurements over a wider range in pseudorapidity and final-state particle multiplicity is important to understand better the origin of these long-range correlations in small collision systems.
In this Letter, measurements of the long-range correlations in \pPb collisions at $\snn = 5.02$~\TeV are extended
to a pseudorapidity gap of $\Delta\eta$$\sim$8 between particles using the ALICE forward multiplicity detectors.
After suppressing non-flow correlations, e.g., from jet and resonance decays, the ridge structure is observed to persist up to a very large gap of $\Delta\eta$$\sim$8 for the first time in \pPb collisions. 
This shows that the collective flow-like correlations extend over an extensive pseudorapidity range also in small collision systems such as \pPb collisions.  
The pseudorapidity dependence of the second-order anisotropic flow coefficient, \Veta, is extracted from the long-range correlations. The \Veta results are presented for a wide pseudorapidity range of $-3.1 <\eta< 4.8$ in various centrality classes in \pPb collisions. To gain a comprehensive understanding of the source of anisotropic flow in small collision systems, the \Veta measurements are compared with hydrodynamic and transport model calculations. The comparison suggests that the final-state interactions play a dominant role in developing the anisotropic flow in small collision systems. 

\end{abstract}
\end{titlepage}

\setcounter{page}{2} 



\section{Introduction}
High-energy heavy-ion collisions can produce a deconfined state of quarks and gluons, the so-called quark--gluon plasma (QGP). Measurements of azimuthal anisotropic flow, for example, via long-range two-particle correlations, are sensitive to key properties of the QGP~\cite{ALICE:2022wpn}. The two-particle correlations between an associated particle and a trigger particle are commonly measured as a function of the differences in pseudorapidity ($\Delta\eta=\eta_{\rm trig}-\eta_{\rm asso}$) and azimuthal angle ($\Delta\varphi=\varphi_{\rm trig}-\varphi_{\rm asso}$). Striking correlations over a long range in $\Delta\eta$ on the near side ($\Delta\varphi\sim0$), the so-called ``ridge'', have been observed in heavy-ion collisions ~\cite{PHOBOS:2008yxa,STAR:2004wfp,PHOBOS:2009sau,STAR:2009ngv,STAR:2011ryj,CMS:2011cqy,ALICE:2011svq,CMS:2012xss,ATLAS:2012at,ALICE:2011ab},  where they are well understood as a consequence of strong, final-state interactions in the dense system created in heavy-ion collisions and the resulting fluid-like collective expansion of the matter created. The correlation function, associate yield as a function of  differences in pseudorapidity and azimuthal angle, projected onto the $\Delta\varphi$ direction, can be expressed in terms of a Fourier series, 
\begin{equation}
\frac{{\rm{d}}N_{\rm pair}}{\rm d\Delta\varphi}\propto1+\sum_{n=1}^{\infty}2v^{2}_{n}\cos{(n\Delta\varphi)},
\end{equation}
where $v_{n}$ is the Fourier coefficient of $n$-th flow harmonics. The $v_{n}$ coefficients 
emerge due to the anisotropic hydrodynamic expansion
of the medium and fluctuate along with the collision geometry in heavy-ion collisions.
The $v_{n}$ coefficients and their multiplicity and transeverse-momentum (\pt) dependence are well described by relativistic hydrodynamic models at midrapidity~\cite{PhysRevD.46.229, PhysRevC.81.054905}.
The pseudorapidity dependence of $v_{n}$ coefficients is sensitive to a temperature dependence of the shear viscosity to entropy density ratio $\eta/s$ of the QGP~\cite{Denicol:2015nhu,Nonaka:2017xsp,Moreland:2018gsh}. It has been measured over a large pseudorapidity region ($|\eta|<5$) in \AuAu and \PbPb  collisions at RHIC~\cite{PHOBOS:2004nvy} and the LHC~\cite{ALICE:2016tlx}.
In small collision systems, such as \pp and \pPb collisions, a ``ridge'' structure was also observed similar to heavy-ion collisions~\cite{PHENIX:2013ktj,PHENIX:2016cfs,PHENIX:2021ubk,STAR:2014qsy,STAR:2022pfn,CMS:2010ifv, CMS:2012qk,CMS:2015fgy,CMS:2016fnw,ALICE:2012eyl,ALICE:2013snk,ATLAS:2012cix,ATLAS:2014qaj}.
The \pt dependence of $v_{n}$ coefficients in small collision systems, which shows a characteristic mass dependence at low \pt,  is found to be similar to heavy-ion collisions~\cite{ATLAS:2014qaj,ALICE:2013snk,CMS:2016fnw,CMS:2018loe}.
Both hydrodynamic (macroscopic) and kinetic transport (microscopic) models  can describe collective-flow observables fairly well in small systems as well as in heavy-ion collisions~\cite{Bozek:2013uha,Weller:2017tsr,Albacete:2013ei,Albacete:2016veq,Zhou:2015iba,Zhao:2017rgg}.
However, the underlying physics of the observed anisotropic flow in small collision systems is still under debate. 
One possible scenario predicts that both contributions from a momentum anisotropy arising in the initial state as well as final interactions are essential in small collision 
systems~\cite{Schenke:2019pmk,Kanakubo:2021qcw,Nie:2019swk}.
Furthermore, the relative contributions of soft collective processes, concentrated in the dense ``core'' and modeled with hydrodynamics, and of hard processes such as hard scattering and jet fragmentation (described as a pp-like ``corona'') are also not well understood.
 The charged-particle pseudorapidity distribution is asymmetric in \pPb collisions~\cite{ALICE:2022imr}: the multiplicity is larger in the Pb-going direction compared to the p-going direction. Since the mean free path depends on the charged-particle multiplicity, the pseudorapidity dependence of \vtwo reflects the underlying dynamical evolution in \pPb collisions and is a direct indicator of how local particle densities modulate the collective flow. 
CMS results on \Veta in \pPb collisions at the LHC~\cite{CMS:2016est,CMS:2017xnj} show a significant pseudorapidity dependence, beyond what would be expected from the pseudorapidity dependence of the mean \pt. 
However, the acceptance is limited to midrapidity $|\eta| < 2$. 
Various small asymmetric collision systems were used to extract 
\Veta at RHIC by means of the event plane method. 
The measurements were carried out within
the pseudorapidity range of $|\eta|<3$~\cite{PHENIX:2018hho,PHENIX:2017nae}. 
A hydrodynamic model~\cite{Bozek:2014cya} describes the result of \Veta qualitatively except for the pseudorapidity regions ($-3<\eta<-2$) in \pAu collisions, where non-flow effects appear to be significant because the rapidity gap from the detector that determines the event plane is small.

In this Letter, the measurements of long-range two-particle correlations and \Veta in \pPb collisions at $\snn = 5.02$~\TeV with the ALICE detector at the LHC are presented. 
These measurements utilize the Forward Multiplicity Detector (FMD) to extract \Veta over the unprecedented range of about 8 units of pseudorapidity ($-3.1<\eta<4.8$). The results are compared with
a hydrodynamic-model calculation and the AMPT transport model.

\section{Experimental setup}
A comprehensive description of the ALICE detector can be found in Refs.~\cite{ALICE:2008ngc,ALICE:2014sbx}. The main detectors used in this analysis are the Forward Multiplicity Detector (\FMD), the Inner Tracking System (\ITS), and the Time Projection Chamber (\TPC).
The \FMD is a silicon strip detector that measures charged particles with a fine granularity of $\Delta\varphi= \pi/20$ and $\Delta\eta=0$.05. 
The \FMD comprises three sub-detectors: \FMDa, \FMDb, and \FMDc. 
The combined acceptance of FMD1 and FMD2 is $1.7<\eta<5.1$. 
The pseudorapidity coverage of FMD3 is $-3.4<\eta<-1.7$.
\FMDa and \FMDc consist of an inner and an outer ring, and \FMDb consists of only one ring, all placed around the beam pipe.
The inner and outer rings are divided into 20 and 40 sectors in the azimuthal direction, respectively, and each ring is composed of hexagonal silicon sensors. Each inner and outer ring is segmented into 512 and 256 strips in the radial direction, respectively. 
Charged-particle tracking at midrapidity is provided using the \TPC and \ITS, both covering the full azimuthal angle and $|\eta|<0.8$. 
They are placed inside the L3 solenoid magnet, which provides a magnetic field of $B=0.5$~T along the beam direction.
The vertex detector, \ITS, consists of six layers of silicon detectors; two layers each are equipped with the Silicon Pixel Detector (\SPD), the Silicon Drift Detector (\SDD), and the Silicon Strip Detector (\SSD).
The \TPC provides track reconstruction
using up to 159 space points along a charged-particle trajectory
and particle identification via the measurement of 
specific energy loss \dEdx.
The \VZERO is used for event triggering and centrality determination.
It is composed of two arrays of 32 scintillator tiles each and covers $-3.7<\eta<-1.7$ (\VZEROC) and $2.8<\eta<5.1$ (\VZEROA). In addition, two neutron Zero Degree Calorimeters (ZDCs) located at $112.5$~m (\ZNA) and $-112.5$~m (\ZNC) from the interaction point along the beam direction are used for the event selection.

\section{Data analysis}
\subsection{Event and Track selection}
This analysis uses ALICE data taken for \pPb collisions at a centre-of-mass energy per nucleon pair of $\snn = 5.02$~\TeV provided by the LHC in 2016, corresponding to a proton beam energy of 4~\TeV and a lead beam energy of 1.58~\TeV per nucleon.
Due to the asymmetric collision system, the nucleon--nucleon centre-of-mass system was shifted by 0.465 rapidity units in the proton beam direction with respect to the ALICE laboratory system. In this Letter, $\eta$ denotes the pseudorapidity in the laboratory system, and the positive pseudorapidity points in the Pb-going direction. 

This analysis uses 5$\times10^{8}$ events acquired using a minimum bias trigger. 
The minimum bias trigger requires the in-time coincidence of signals from \VZEROA and \VZEROC. 
The first step of the event selection is performed on the amplitude and timing in the \VZERO and \ZDC as described in Ref.~\cite{ALICE:2012xs}. 
The efficiency of the event selection is 99.2$\%$ for non-single-diffractive collisions. 
The primary vertex position is determined with reconstructed tracks in the \TPC and \ITS by using an analytic $\chi^{2}$ minimization method as described in Ref.~\cite{ALICE:2012aqc}. 
The primary vertex in the beam axis is required to be within 10 cm of the nominal interaction point along the beam line.
In addition, pile-up events from beam-induced background are rejected by using correlations between the \FMD and \VZERO multiplicities. 

Charged-particle tracks are reconstructed in the ITS and TPC within $|\eta|<0.8$ for $\pt>0.2$~\GeVc as follows.
First, the tracks are selected on the number of space points and the quality of the track fit in the \TPC. In addition, the tracks are required to have a distance of closest approach (\DCA) to the reconstructed primary vertices less than 2~cm in the 
beam-axis direction. 
The \DCA in the transverse direction is required to be less than 7~$\sigma_{\rm DCA}$, where $\sigma_{\rm DCA}$ is a \pt-dependent transverse impact parameter resolution.
The efficiency of the charged-particle track selection is estimated with a Monte Carlo (MC) simulation using the DPMJET event generator~\cite{Roesler:2000he} and GEANT3~\cite{Brun:1082634} to simulate particle transport through the detector. The efficiency is about 65\% at \pt=0.2 \GeVc, increases to about 79\% at \pt=0.8~\GeVc, and decreases to about 76\% at \pt=3 \GeVc.

The selected events are divided into several centrality classes based on the \VZEROA amplitude. Table~\ref{tab:eventclass} shows the event classes and corresponding average charged-particle pseudorapidity densities within $-3.25<\eta<-2.5$, $|\eta|<0.8$, $2<\eta<3.75$, and $3.75<\eta<5$. The multiplicities were measured by the innermost two layers of the \ITS at midrapidity and the \FMD at forward and backward rapidity~\cite{ALICE:2022imr}.

\begin{table}[htbp]
  \begin{center}
    \caption{Average charged-particle pseudorapidity density for $\pt>0$~\GeVc in different pseudorapidity regions.}
      \begin{tabular}{ccccc} \hline
         Centrality & $\langle\dndeta \rangle_{\rm -3.25<\eta<\rm -2.5}$ & $ \langle\dndeta \rangle_{|\eta|<\rm 0.8}$ & $\langle\dndeta\rangle_{\rm 2<\eta<\rm 3.75}$ & $\langle\dndeta\rangle_{\rm 3.75<\eta<\rm 5}$\\ \hline
        
        0--5$\%$   & $34\pm2.2$   & $45\pm1.4$  & $60\pm3.8$    &$52\pm3.5 $ \\
        5--10$\%$  & $29\pm1.9$   & $36\pm1.1$    & $46\pm2.9$    &$39\pm2.6  $  \\
        10--20$\%$ & $26\pm1.6$   & $31\pm0.94$    & $37\pm2.4$   &$31\pm2.1  $  \\
        20--40$\%$ & $21\pm1.4$   & $23\pm0.72$  & $27\pm1.7$    &$22\pm1.5 $ \\
        40--60$\%$ & $16\pm1.0$   & $16\pm0.54$  & $17\pm1.1$   &$14\pm0.95 $ \\ 
        60--80$\%$ & $11\pm0.68$ & $9.7\pm0.33$ & $9.9\pm0.63$ &$7.8\pm0.52$  \\
        80--100$\%$& $5.4\pm0.34$ & $4.2\pm0.14$ & $3.5\pm0.22$ &$2.7\pm0.18$  \\\hline
      \end{tabular}
    \label{tab:eventclass}
  \end{center}
\end{table}%

\subsection{Two-particle correlation and extraction of \texorpdfstring{$v_{2}(\eta)$}{}}
\label{sec:analysis}
For a given event class, two-particle correlations between a trigger and associated particle are measured as a function of the pseudorapidity difference $\Delta\eta$ and the azimuthal angle difference $\Delta\varphi$. 
The associated yield to a trigger particle as a function of $\Delta\eta$ and $\Delta\varphi$ is defined as
\begin{equation}
    \frac{1}{N_{\rm trig}}\frac{\rm d^{2} \it N_{\rm assoc}}{\rm d\Delta\eta d\Delta\varphi}=\frac{S(\Delta\eta,\Delta\varphi)}{B(\Delta\eta,\Delta\varphi)},
\end{equation}
where $N_{\rm trig}$ is the total number of trigger particles in the given event class, the signal function $S(\Delta\eta,\Delta\varphi) =\frac{1}{\it N_{\rm  trig}}\frac{ {\rm d^{2}}N_{\rm same}}{\rm d\Delta\eta \rm d\Delta\varphi}$ is the associated yield per trigger particle in the same event, and the background function $B(\Delta\eta,\Delta\varphi) =\alpha\frac{\rm d^{2}\it N_{\rm mixed}}{\rm d\Delta\eta \rm d\Delta\varphi}$ is the pair yield 
associated to a trigger particle when the associated particles are taken from other events which fall in the same event class. 
The $\alpha$ factor is chosen such that $B(\Delta\eta,\Delta\varphi)$ is unity at its maximum.
By dividing $S(\Delta\eta,\Delta\varphi)$ by $B(\Delta\eta,\Delta\varphi)$, pair acceptance and single particle efficiency for both particles are corrected. 
To take into account the vertex position dependence of the above acceptance and efficiency, the correlation function is extracted 
in 2 cm wide intervals of the vertex position. 

\begin{figure}[tb]
    \begin{center}
    \includegraphics[width= \textwidth]{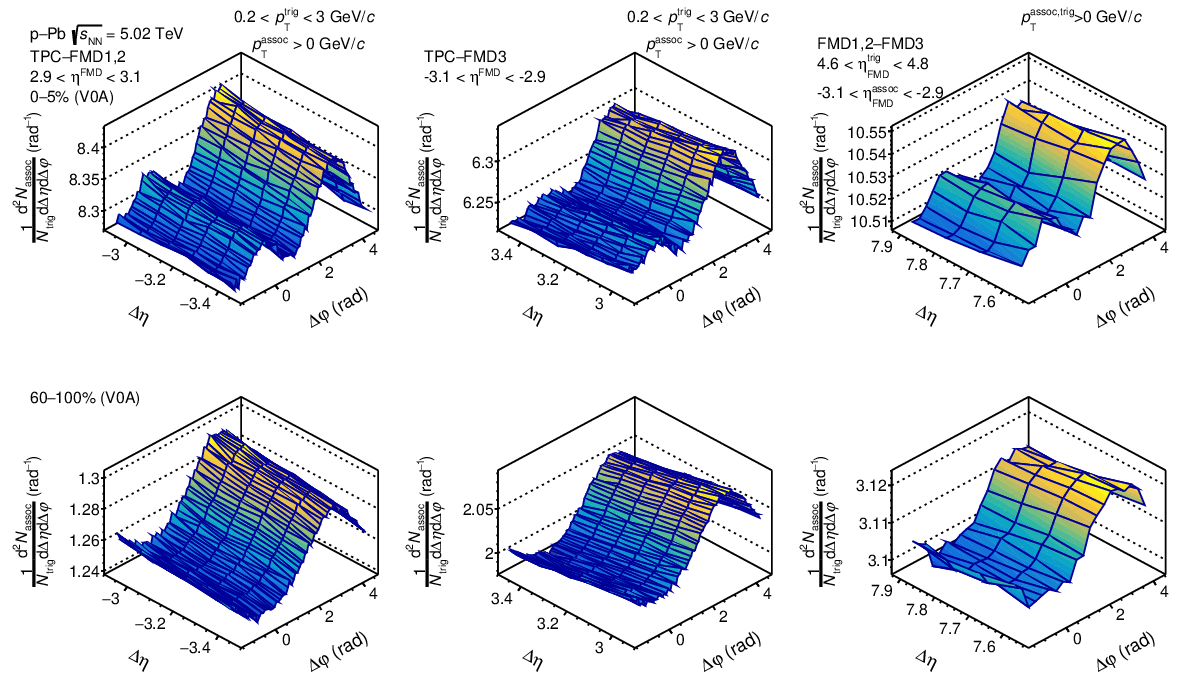}
    \end{center}
    \caption{The associated yield per trigger as a function of $\Delta\eta$ and $\Delta\varphi$ as measured for \TPC--\FMDab (left), \TPC--\FMDc (central), and \FMDab--\FMDc (right) correlations in the 0--5\% (top) and 60--100\% (bottom) \pPb collisions.}
    \label{fig:2d}
\end{figure}

Figure~\ref{fig:2d} shows the correlation function in $\Delta\eta$ and $\Delta\varphi$ between trigger and associated particles in 
\TPC ($|\eta|<0.8$) -- \FMDab ($2.9<\eta<3.1$) (left), 
\TPC -- \FMDc ($-3.1<\eta<-2.9$) (center), 
and \FMDab ($4.6<\eta<4.8$) -- \FMDc ($-3.1<\eta<-2.9$) (right) 
in the 0--5\% (top) and 60--100$\%$ (bottom) 
 \pPb collisions at \snn = 5.02~\TeV, respectively. 
For all three combinations, a long-range correlation on the near side ($-\pi/2<\Delta\varphi<\pi/2$), the so-called ``ridge'', is observed in the 0--5\% event class, while no significant ``ridge'' is observed in 60--100\%. The long-range correlation in the away side ($\pi/2<\Delta\varphi<3\pi/2$) mainly results from jets recoiling opposite to the trigger particle and is visible in all event classes.
Since the pseudorapidity gap between the trigger and associated particles is large,  the pronounced peak structure on the near side due to jets~\cite{ALICE:2012eyl}, which is centred on $\Delta\eta=0$, is not present.
In central \pPb collisions at the LHC, the near-side ``ridge'' structure is observed to extend up to a pseudorapidity separation of $\Delta\eta\sim8$, which is the largest range measured.

To estimate and subtract the non-flow effects due to recoil jets and resonance decays, the template fit procedure, which was introduced by the ATLAS collaboration, is employed~\cite{ATLAS:2015hzw,ATLAS:2016yzd}.
The correlation function $Y(\Delta\varphi)$ is assumed to be a superposition of a non-flow contribution, which is estimated by scaling the correlation function from peripheral events, and the flow contribution.
The template fit function is defined as
\begin{equation}
\label{eqtemplate}
Y(\Delta\varphi)=FY^{\rm peri}(\Delta\varphi)+G^{\rm tmp}\lbrace 1+2\sum_{n=2}^{3}V^{\rm tmp}_{n,n}\cos(n\Delta\varphi)\rbrace,
\end{equation}
where $F, G^{\rm tmp}$, and $V^{\rm tmp}_{n,n}$ are free parameters. 
$Y^{\rm peri}(\Delta\varphi)$ is the correlation function in peripheral events. 
$V^{\rm tmp}_{n,n}$ represents the $n$-th flow coefficient, and $F$ and $G$ are the scaling factors of the non-flow distribution on the away side and a flat, azimuth-independent baseline, respectively.
This method assumes no flow components in the peripheral event used for the template and no away-side jet modifications between central and peripheral events.

\begin{figure}[tb]
    \begin{center}
    \includegraphics[width= \textwidth]{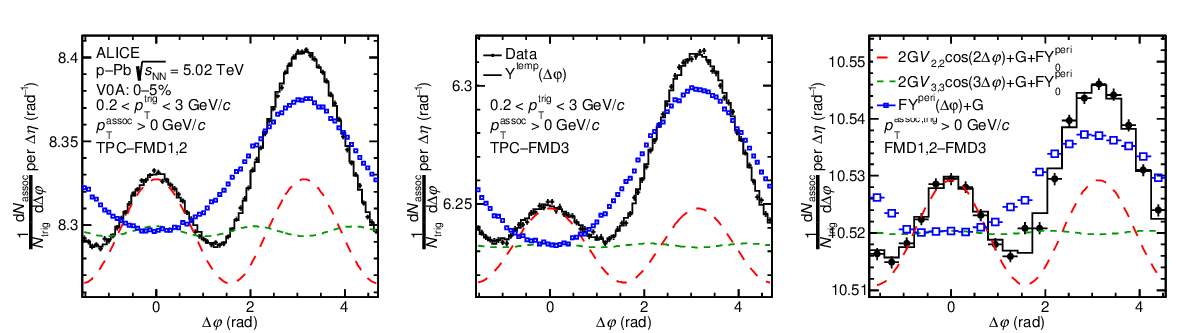}
    \end{center}
    \caption{Projection of the correlation function of \TPC--\FMDab (left), \TPC--\FMDc (central), and \FMDab--\FMDc (right) correlations in 0--5\% \pPb collisions  with the template fit using Eq.~\eqref{eqtemplate}.
     The open circle blue marker represents the scaled peripheral distribution plus the Flow baseline, $G$. The red and green dashed lines represent the second- and third-order components plus the baseline, respectively.}
      
    \label{fig:pro}
\end{figure}

Figure~\ref{fig:pro} shows the projection of the correlation functions, i.e., \TPC--\FMDab (left), \TPC--\FMDc (center), \FMDab--\FMDc (right), onto the $\Delta\varphi$ axis in the 0--5\% event class. These one-dimensional correlations are fitted by the template fit function using Eq.~\eqref{eqtemplate}. 
The fit describes the data well with a $\chi^{2}$/ndf of about 0.8--2.5 for all correlation functions corresponding to all $\eta$ gap combinations. 
The flow-like part of the fit is dominated by the second harmonic 
for all three correlations. 
The modulation for the pair correlation is extracted for each harmonic in the template fit.

Assuming that the relative modulation of the two-particle correlation function is solely due to the modulation of the single-particle distribution, the modulation of the two-particle correlation measured in two different pseudorapidity ranges for particles a and b can be factorized as:
\begin{equation}
\label{eqfactorization}
    V_{n,n}(\eta_{a},\eta_{b})=v_{n}(\eta_{a})v_{n}(\eta_{b}).
\end{equation}
If this factorization holds, the flow component of a single particle at a certain pseudorapidity can be extracted from three dihadron correlations between different pseudorapidity regions given by
\begin{equation}
\label{eq:threesub}
    v_{n}(\eta_{a})=\sqrt{\frac{V_{n,n}(\eta_{a},\eta_{b})V_{n,n}(\eta_{a},\eta_{c})}{V_{n,n}(\eta_{b},\eta_{c})}},
\end{equation}
where each $V_{n,n}$ is the modulation extracted by TPC--FMD1,2, TPC--FMD3, and FMD1,2--FMD3 correlation, respectevily.
This method is similar to the ``3$\times$2PC'' method used by PHENIX~\cite{PHENIX:2021ubk,PHENIX:2022nht}.
The factorization breaks down if the event plane 
and/or the flow amplitude depend on pseudorapidity, for example because of initial longitudinal fluctuations or thermal fluctuations~\cite{ATLAS:2017rij,ATLAS:2020sgl,CMS:2015xmx,Sakai:2021rug,Bozek:2017qir,Bozek:2015bna}. Therefore, the uncertainty due to those decorrelation effects is estimated by changing the $\eta$ gap, 
as it will be discussed in Section~\ref{sec:systematic}.

Since the \FMD is not a tracking detector, it is difficult to separate primary particles from secondary particles inside the \FMD. 
Secondary particles are generated around the primary particle and might distort its distribution.
The effects of secondary particles on the flow harmonics are estimated using MC simulations based on the AMPT and EPOS event generators~\cite{Lin:2004en,Albacete:2016veq,Pierog:2013ria}. The correction is performed based on the change in the reconstructed particle distribution after particle transport and interaction within the detector material from the original distribution.
The correction factor is extracted as the ratio of the \vtwo of primary particles over the  \vtwo of all reconstructed particles (i.e. including primary and secondary particles). 
In order to match the range of the FMD, which has acceptance down to $\pt=0$, the charged-particle \vtwo at midrapidity is extrapolated to $\pt=0$ based on the data of the \pt spectrum and the \pt differential \vtwo of charged-particles. 
The factor is about 0.86 for all four centralities.

\subsection{Systematic uncertainties}
\label{sec:systematic}
The systematic uncertainties relate to the event selection, the track selection, the correction of secondary particles in the \FMD, the material budget in the \FMD, the choice of the peripheral event class for non-flow subtraction, the reference pseudorapidity choice of $\eta_{\rm b},\eta_{\rm c}$ to extract $v_2(\eta_{\rm a})$, and the jet modification.
A systematic uncertainty is only assigned when the difference between the nominal data points and variations is statistically significant according to the Barlow criterium~\cite{Barlow:2002yb}.
Table~\ref{tab:systematic} shows the summary of systematic uncertainties for \vtwo,
 which depend on centrality and pseudorapidity as indicated by the range given.
The uncertainty due to the event selection is investigated by changing  the selection parameters for \VZERO--\FMD multiplicity correlations. 
The uncertainty slightly depends on centrality, and it is the largest in the 20--40\% centrality class.
The uncertainty due to track selection is evaluated by varying track selection parameters. This systematic uncertainty is 0.28--0.45\% 
and is also the largest in the 20--40\% centrality class.
The systematic uncertainty related to the correction of the contamination by secondary particles is estimated using different event generators, EPOS and AMPT.
The magnitude of collective-like signal in EPOS and AMPT is very different, and this check investigates how stable the correction is with respect to the magnitude of the flow. This uncertainty is found to be larger in the p-going direction than in the Pb-going direction.
The number of secondary particles depends on the material budget in the ALICE environment. This systematic uncertainty is 2.3$\%$ estimated using MC simulations with increased or reduced material budget of the detector descriptions in GEANT simulation by $\pm$10\%.  
Uncertainties associated with secondary correction and material budget are assigned only for the \FMD acceptance.
The systematic uncertainty of the peripheral event choice used for non-flow subtraction is estimated using the 80--100$\%$ event class instead of 60--100$\%$. 
The systematic uncertainty on the reference pseudorapidity choice is investigated by changing the $\eta$ gap between $\eta_{\rm b}$ and $\eta_{\rm c}$ to extract $v_2(\eta_{\rm a})$.
The uncertainty ranges from 2.6$\%$ to 5.1$\%$ depending on the centrality and the choice of reference pseudorapidity.
The shape of the jet is modified depending on the centrality. 
The uncertainty is evaluated by varying the away-side width of the peripheral collisions and is 0.8--2.7\%.
These systematic uncertainties are added in quadrature.

\begin{table}[htbp]
  \begin{center}
    \caption{The summary of systematic uncertainties, which is absolute value on \vtwo. The uncertainties take values within the given range depending on the centrality and pseudorapidity for each source.}
      \begin{tabular}{c|c} \hline
        Source of uncertainty         & Systematic uncertainty (\%)\\\hline
        Event selection & 0.56--3.1$\%$ \\
        Track selection & 0.28--0.45$\%$\\
        Secondary correction & 0.74--3.5$\%$\\
        Material budget & 2.3$\%$\\
        Non-flow sub  & 0.76--4.6$\%$\\
        Jet modification & 0.8--2.7\%\\
        $\eta$ gap selection & 2.6--5.1$\%$\\
        \hline
        Total & 3.9--8.3$\%$\\\hline
      \end{tabular}
    \label{tab:systematic}
  \end{center}
\end{table}

\section{Results}
\begin{figure}[]
 \hspace{1cm}
    \includegraphics[width= 0.75\textwidth]{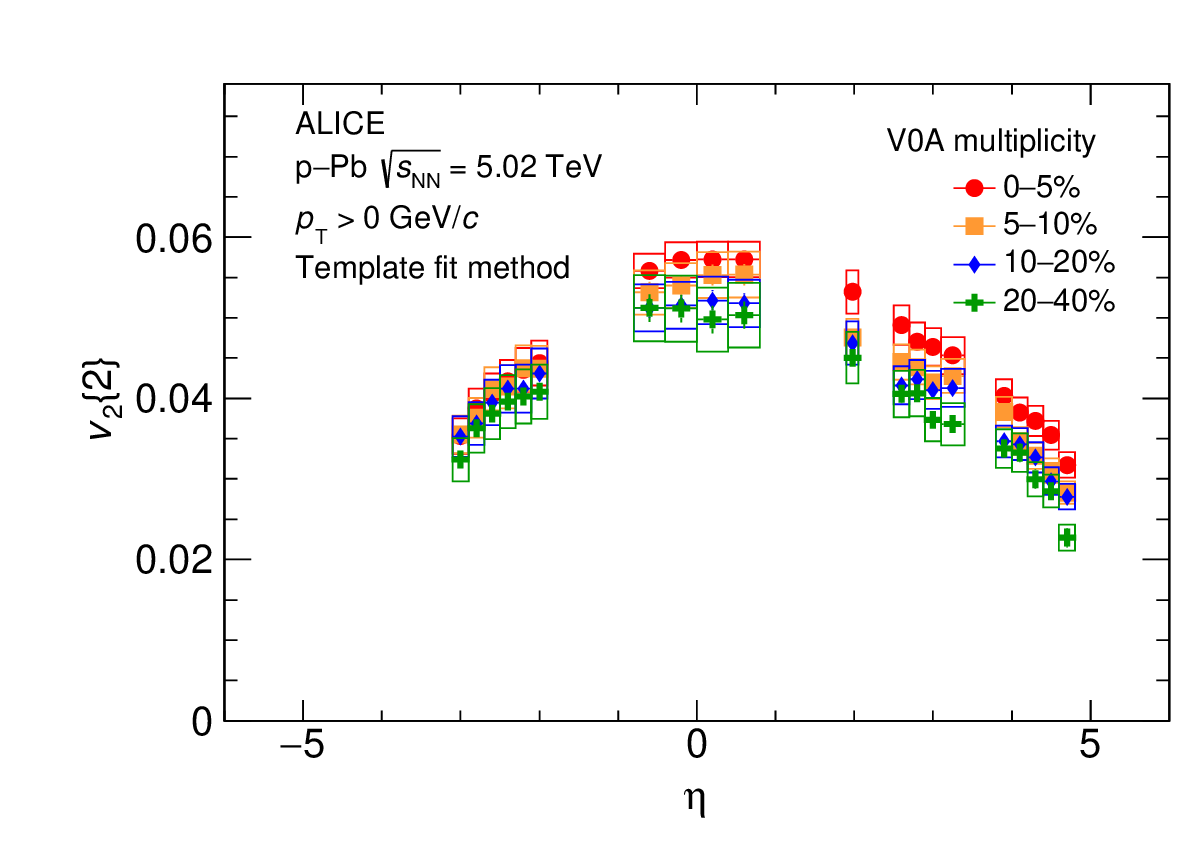}
    \caption{$p_{\mathrm T}$-integrated $v_{2}\{2\}$ as a function of $\eta$ in various centrality classes using the template fitting method. Boxes show the total systematic uncertainties.}
    \label{fig:template}
\end{figure}

The \vtwo for charged-particles in a specific pseudorapidity region can be extracted using the template fit procedure and the three dihadron correlations of TPC--FMD1,2, TPC--FMD3, and FMD1,2--FMD3, as described in Section~\ref{sec:analysis}. 
The corresponding results of the \pt-integrated \vtwo as a function of $\eta$ for the 0--5\%, 5--10\%, 10--20\%, and 20--40\% centrality classes are shown in Fig.~\ref{fig:template}.
After applying the non-flow subtraction with the template fit approach, a non-zero \vtwo is observed over a wide rapidity range for the first time in \pPb collisions. The \vtwo results in \pPb collisions were measured previously for the pseudorapidity range of $|\eta|<2$ by CMS~\cite{CMS:2016est, CMS:2017xnj}. The \vtwo measurements presented in this Letter significantly extend the measurements to a much wider pseudorapidity range, $-3.1<\eta<4.8$. This result confirms the emergence of anisotropic flow over a wide rapidity region, as in high-energy heavy-ion collisions~\cite{ALICE:2016tlx}. In addition, the \vtwo measurements show a significant pseudorapidity dependence for all four centrality classes. It is more prominent in the Pb-going direction (positive $\eta$) than in the p-going direction (negative $\eta$). Additionally, a stronger centrality dependence of \vtwo is found in the Pb-going direction than in the p-going direction.

\begin{figure}[]
\hspace{1cm}
    \includegraphics[width= 0.75\textwidth]{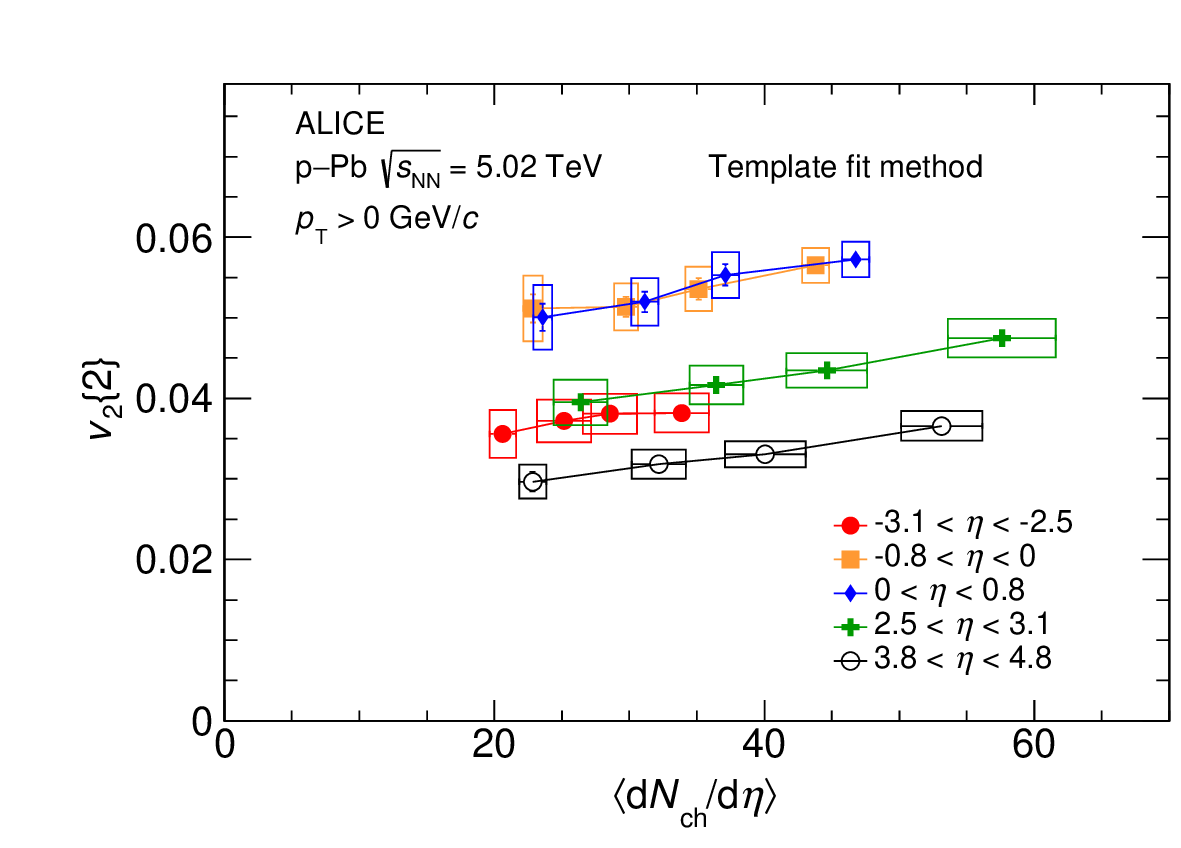}
    \caption{\vtwo as a function of charged-particle pseudorapidity density for five different pseudorapidity regions. }
    
    \label{fig:dndeta}
\end{figure}

Previous \Veta  measurements showed that the magnitude of \vtwo is correlated with charged-particle pseudorapidity density~\cite{PHENIX:2018hho}; \Veta increases with increasing multiplicity. However, \Veta might not be linearly correlated with \dndeta~\cite{ALICE:2022imr}. 
Figure~\ref{fig:dndeta} shows \vtwo as a function of charged-particle multiplicity density for five different pseudorapidity regions 
and for different centrality classes: 0--5\%, 5--10\%, 10--20\%, 
and 20--40\%.
Figure~\ref{fig:dndeta} demonstrates that the pseudorapidity dependence of \vtwo is not just simply driven by the local multiplicity, \vtwo independently depends on both $\eta$ and \dndeta. In a fixed pseudorapidity range, \vtwo depends on local multiplicity, but at fixed local multiplicity, there is still a significant dependence on the pseudorapidity.

\begin{figure}[]
\hspace{1cm}
    \includegraphics[width= 0.75\textwidth]{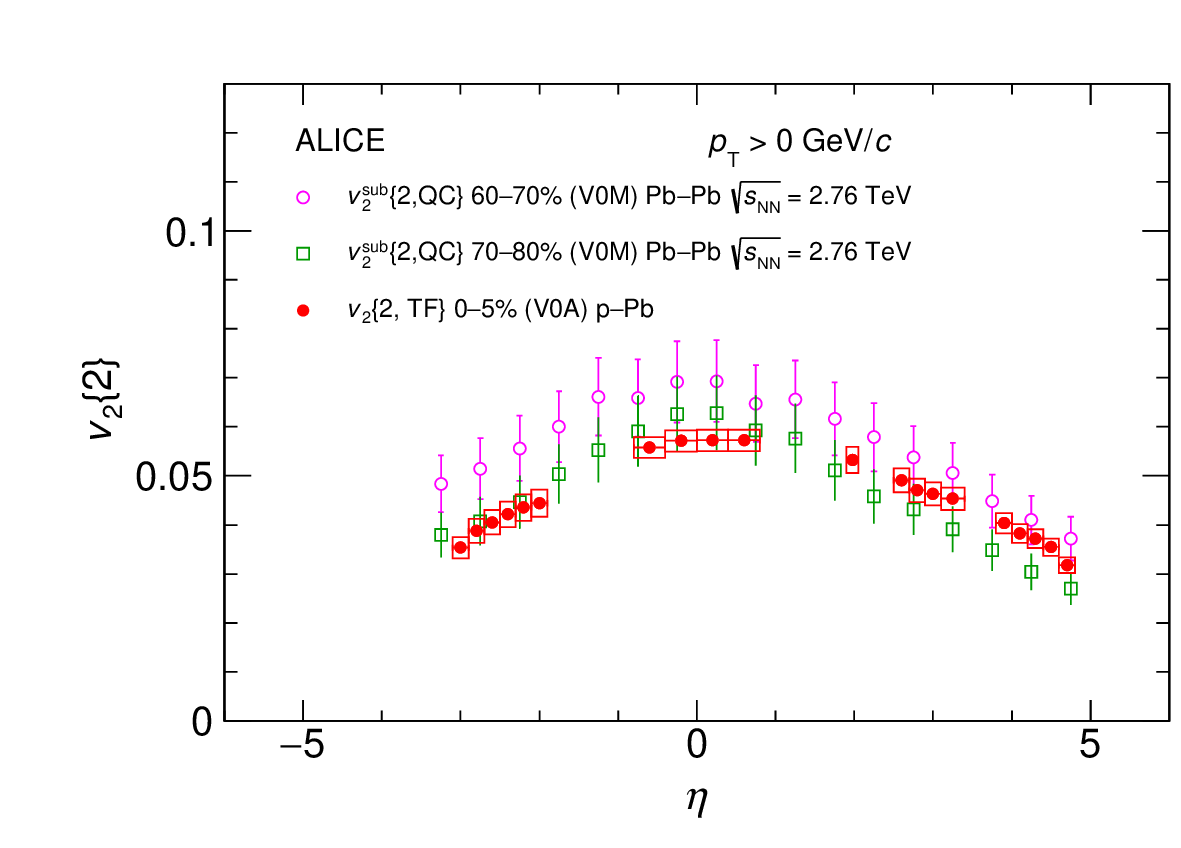}
    \caption{The \Veta in central \pPb collisions compared with \Veta in peripheral \PbPb collisions with a compatible mean charged-particle multiplicity in Pb-going direction. 
    The \vtwo \PbPb results were obtained using the Q-cumulant method~\cite{ALICE:2016tlx}.}
   
    \label{fig:compPbPb}
\end{figure}

Figure~\ref{fig:compPbPb} shows the comparisons of \vtwo in 0--5\% \pPb collisions and in 60--70\% and 70--80\% \PbPb collisions, where the multiplicity is similar in the Pb-going direction between the two collision systems~\cite{ALICE:2022imr,ALICE:2015bpk}. The charged-particle pseudorapidity density at $\eta\sim3$ is about 60 in \pPb collisions and 80 and 37 in 60--70\% and 70--80\% \PbPb collisions, respectively.
Here, the results from \PbPb collisions are based on the 2-particle cumulant method~\cite{ALICE:2016tlx}. The \vtwo results in \pPb collisions are compatible with the \vtwo in 60--70\% and 70--80\% \PbPb collisions over the entire $\eta$ range within the sizable uncertainties of the \PbPb results. 
Somewhat unexpectedly, given the potential differences in the initial collision overlap geometry, it is observed that peripheral \PbPb collisions and central \pPb collisions have comparable \vtwo at similar multiplicities.

\begin{figure}[]
    \begin{center}
\includegraphics[width= 0.75\textwidth]{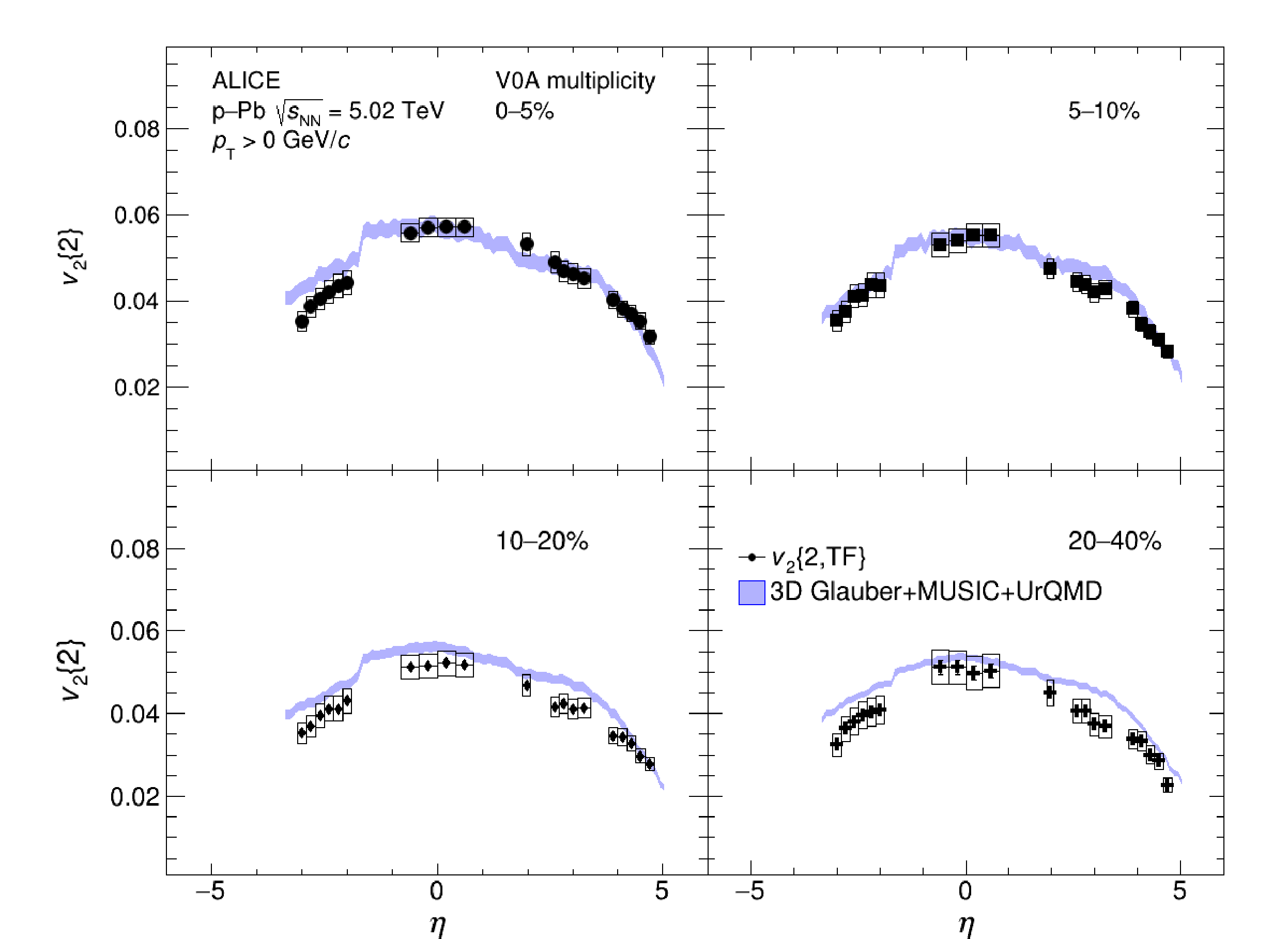}
    \end{center}
    \caption{
    Pseudorapidity dependence of \pt integrated \vtwo. 
    Comparison of the 
    measured data (black circles) with a calculation by the hydrodynamical model (blue band)~\cite{Zhao:2022ayk} for the 0--5\% (top left), 5--10\% (top right), 10--20\% (bottom left), and 20--40\% (bottom right) centrality classes.}
   
    \label{fig:hydro}
\end{figure}

\begin{figure}[]
    \begin{center}
    \includegraphics[width= 0.75\textwidth]{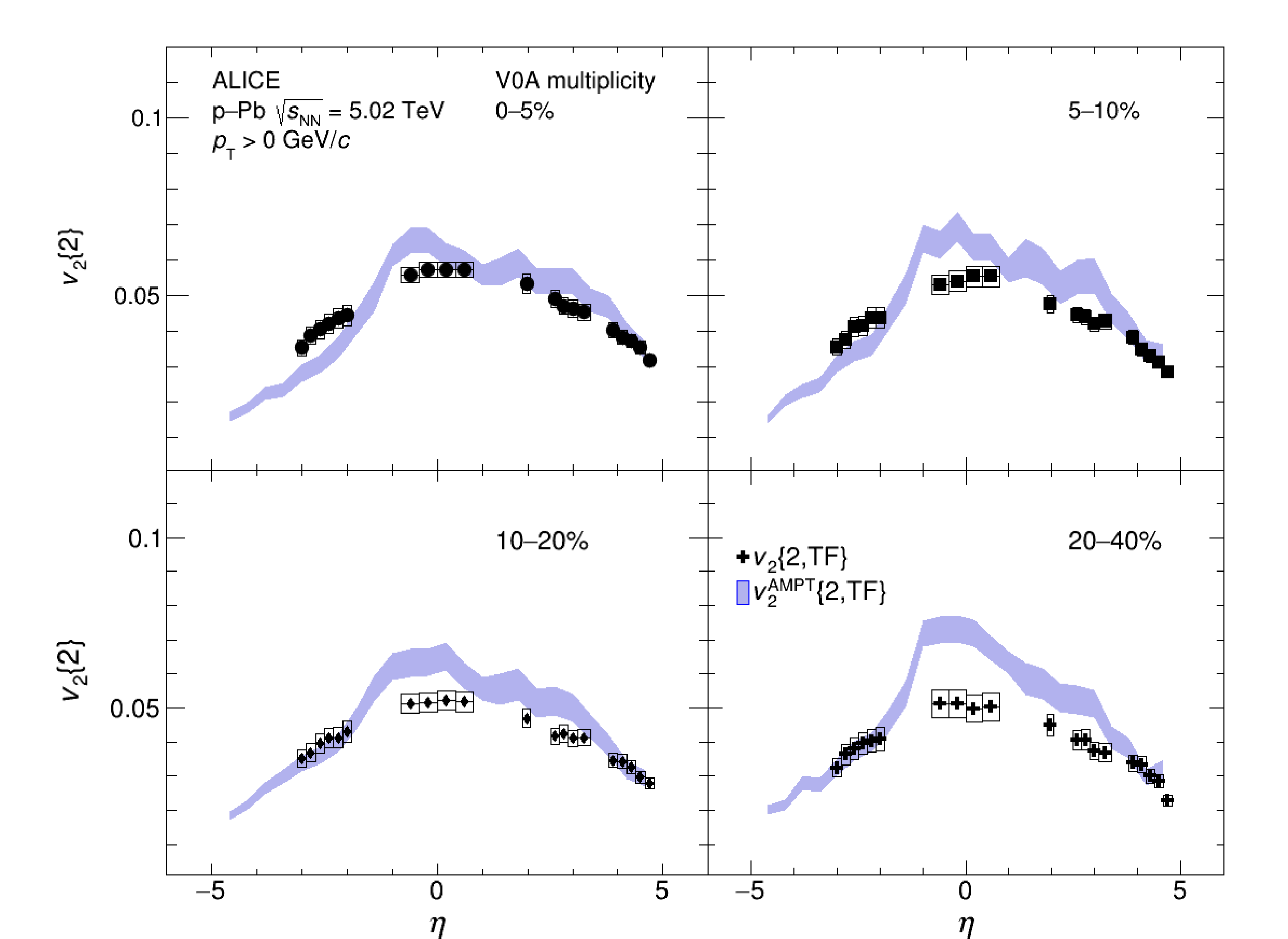}
    \end{center}
    \caption{
    Pseudorapidity dependence of \pt integrated \vtwo. 
    Comparison of the 
    measured data  (black circles) with a calculation by AMPT with the string-melting configuration (blue band) for 0--5\% (top left), 5--10\% (top right), 10--20\% (bottom left), and 20--40\% (bottom right) centrality class.}
    \label{fig:amptimp}
\end{figure}

To further investigate the origin of the flow in small collision systems, the \Veta measured in \pPb collisions is compared with hydrodynamical calculations~\cite{Zhao:2022ayk} in Fig.~\ref{fig:hydro}, and with the AMPT transport model in  Fig.~\ref{fig:amptimp}. 
The 3+1 hydrodynamical model employs 3D Glauber initial conditions, viscous hydrodynamics based on MUSIC, and the UrQMD model to simulate the dynamics in the hadronic phase. 
The shear viscosity and color string width in the transverse plane are adjusted to $\eta_{T}/(e+P)=0.08$ and $\sigma_{x}=0.4$~fm, respectively, to reproduce the mean transverse momentum of identified particles and the \pt dependence of charged-particle \vtwo measured with the template fit procedure by ATLAS at midrapidity in \pPb collisions at $\snn=5.02$~\TeV~\cite{Zhao:2022ayk}.  
The hydrodynamical model underestimates the pseudorapidity dependence of charged-particle multiplicity density when centrality is determined at forward pseudorapidity, while it is reproduced when centrality is determined at midrapidity ~\cite{Shen:2022oyg}.
The correlation between the multiplicities at forward and midrapidity is weaker in the model than in the data.
In this model, \vtwo mainly originates from the 3D initial geometry and develops in the course of the hydrodynamical evolution.
The model describes the \Veta measurement in 0--5\% and 5--10\%, 
while it somewhat overestimates the data in 10--20\% and 20--40\% at both forward and backward rapidity.

Similarly, Fig.~\ref{fig:amptimp} shows the comparisons with calculations by the AMPT model in the string-melting configuration.
Unlike the hydrodynamical model, the AMPT with string melting is a transport model that produces collective behaviour microscopically by final-state scattering in both the partonic and hadronic phases. 
The \Veta calculation from AMPT describes the data qualitatively in the 0--5\% centrality class and reproduces the asymmetry between the Pb-going and p-going directions. However, the centrality dependence is less significant than observed in the data.  

The comparisons between the \Veta measurements and the calculations from the hydrodynamical and AMPT transport models suggest that strong 
final-state interactions are possibly the origin of a significant \vtwo over a wide pseudorapidity range in small collision systems such as \pPb collisions. 
Initial momentum anisotropy from Colour-Glass Condensate (CGC) could also play a role in high-multiplicity \pPb collisions. However, a recent study using gluon saturation from the IP-Glasma model shows that the initial momentum anisotropy results to short-range correlation~\cite{Schenke:2022mjv}. The observed finite \vtwo extracted from ultra-long correlations will likely not be influenced by contributions from initial momentum anisotropy from CGC but more likely originates from the fluctuating initial geometry giving rise to final-state interactions. 

\section{Summary}
In this Letter, the two-particle correlation function is presented as a function of $\Delta\eta$ and $\Delta\varphi$ in \pPb collisions at $\snn = 5.02$~\TeV with ALICE. The ``ridge'' structure is observed up to a rapidity gap of 8 units between the trigger and the associate particles,  in central events. A double-ridge structure is visible after the non-flow subtraction. The \pt-integrated \vtwo is extracted from the correlation function and presented as a function of pseudorapidity and centrality class. Non-zero \vtwo is observed over a wide pseudorapidity range in central \pPb collisions for the first time. 
The pseudorapidity dependence as well as its asymmetric shape of \vtwo could be well explained by charged-particle multiplicity distributions. 
In addition, the \Veta in \pPb central events is comparable with the \Veta in peripheral \PbPb collisions at similar multiplicity. 
Finally, both hydrodynamical~\cite{Zhao:2022ayk} and AMPT transport model calculations~\cite{Albacete:2016veq}
describe the data qualitatively over a wide pseudorapidity region.
The comparison with the model calculations suggests the emergence of collective flow at very forward pseudorapidity ($|\eta|\sim5$) region in \pPb collisions, just like in high-energy heavy-ion collisions. 
The results suggest an important role of final-state interactions in developing anisotropic flow in small collision systems.


\newenvironment{acknowledgement}{\relax}{\relax}
\begin{acknowledgement}
\section*{Acknowledgements}

The ALICE Collaboration would like to thank all its engineers and technicians for their invaluable contributions to the construction of the experiment and the CERN accelerator teams for the outstanding performance of the LHC complex.
The ALICE Collaboration gratefully acknowledges the resources and support provided by all Grid centres and the Worldwide LHC Computing Grid (WLCG) collaboration.
The ALICE Collaboration acknowledges the following funding agencies for their support in building and running the ALICE detector:
A. I. Alikhanyan National Science Laboratory (Yerevan Physics Institute) Foundation (ANSL), State Committee of Science and World Federation of Scientists (WFS), Armenia;
Austrian Academy of Sciences, Austrian Science Fund (FWF): [M 2467-N36] and Nationalstiftung f\"{u}r Forschung, Technologie und Entwicklung, Austria;
Ministry of Communications and High Technologies, National Nuclear Research Center, Azerbaijan;
Conselho Nacional de Desenvolvimento Cient\'{\i}fico e Tecnol\'{o}gico (CNPq), Financiadora de Estudos e Projetos (Finep), Funda\c{c}\~{a}o de Amparo \`{a} Pesquisa do Estado de S\~{a}o Paulo (FAPESP) and Universidade Federal do Rio Grande do Sul (UFRGS), Brazil;
Bulgarian Ministry of Education and Science, within the National Roadmap for Research Infrastructures 2020--2027 (object CERN), Bulgaria;
Ministry of Education of China (MOEC) , Ministry of Science \& Technology of China (MSTC) and National Natural Science Foundation of China (NSFC), China;
Ministry of Science and Education and Croatian Science Foundation, Croatia;
Centro de Aplicaciones Tecnol\'{o}gicas y Desarrollo Nuclear (CEADEN), Cubaenerg\'{\i}a, Cuba;
Ministry of Education, Youth and Sports of the Czech Republic, Czech Republic;
The Danish Council for Independent Research | Natural Sciences, the VILLUM FONDEN and Danish National Research Foundation (DNRF), Denmark;
Helsinki Institute of Physics (HIP), Finland;
Commissariat \`{a} l'Energie Atomique (CEA) and Institut National de Physique Nucl\'{e}aire et de Physique des Particules (IN2P3) and Centre National de la Recherche Scientifique (CNRS), France;
Bundesministerium f\"{u}r Bildung und Forschung (BMBF) and GSI Helmholtzzentrum f\"{u}r Schwerionenforschung GmbH, Germany;
General Secretariat for Research and Technology, Ministry of Education, Research and Religions, Greece;
National Research, Development and Innovation Office, Hungary;
Department of Atomic Energy Government of India (DAE), Department of Science and Technology, Government of India (DST), University Grants Commission, Government of India (UGC) and Council of Scientific and Industrial Research (CSIR), India;
National Research and Innovation Agency - BRIN, Indonesia;
Istituto Nazionale di Fisica Nucleare (INFN), Italy;
Japanese Ministry of Education, Culture, Sports, Science and Technology (MEXT) and Japan Society for the Promotion of Science (JSPS) KAKENHI, Japan;
Consejo Nacional de Ciencia (CONACYT) y Tecnolog\'{i}a, through Fondo de Cooperaci\'{o}n Internacional en Ciencia y Tecnolog\'{i}a (FONCICYT) and Direcci\'{o}n General de Asuntos del Personal Academico (DGAPA), Mexico;
Nederlandse Organisatie voor Wetenschappelijk Onderzoek (NWO), Netherlands;
The Research Council of Norway, Norway;
Commission on Science and Technology for Sustainable Development in the South (COMSATS), Pakistan;
Pontificia Universidad Cat\'{o}lica del Per\'{u}, Peru;
Ministry of Education and Science, National Science Centre and WUT ID-UB, Poland;
Korea Institute of Science and Technology Information and National Research Foundation of Korea (NRF), Republic of Korea;
Ministry of Education and Scientific Research, Institute of Atomic Physics, Ministry of Research and Innovation and Institute of Atomic Physics and University Politehnica of Bucharest, Romania;
Ministry of Education, Science, Research and Sport of the Slovak Republic, Slovakia;
National Research Foundation of South Africa, South Africa;
Swedish Research Council (VR) and Knut \& Alice Wallenberg Foundation (KAW), Sweden;
European Organization for Nuclear Research, Switzerland;
Suranaree University of Technology (SUT), National Science and Technology Development Agency (NSTDA) and National Science, Research and Innovation Fund (NSRF via PMU-B B05F650021), Thailand;
Turkish Energy, Nuclear and Mineral Research Agency (TENMAK), Turkey;
National Academy of  Sciences of Ukraine, Ukraine;
Science and Technology Facilities Council (STFC), United Kingdom;
National Science Foundation of the United States of America (NSF) and United States Department of Energy, Office of Nuclear Physics (DOE NP), United States of America.
In addition, individual groups or members have received support from:
European Research Council, Strong 2020 - Horizon 2020 (grant nos. 950692, 824093), European Union;
Academy of Finland (Center of Excellence in Quark Matter) (grant nos. 346327, 346328), Finland.

\end{acknowledgement}

\bibliographystyle{utphys}   
\bibliography{bib/introduction,bib/bibliography,bib/setup,bib/event,bib/analysis,bib/systematic,bib/bibliography_cms,bib/result}

\providecommand{\href}[2]{#2}\begingroup\raggedright\begin{thebibliography}{10}

\bibitem{ALICE:2022wpn}
{\bfseries ALICE} Collaboration, ``{The ALICE experiment -- A journey through
  QCD}'', \href{http://arxiv.org/abs/2211.04384}{{\ttfamily arXiv:2211.04384
  [nucl-ex]}}.

\bibitem{PHOBOS:2008yxa}
{\bfseries PHOBOS} Collaboration, B.~Alver {\em et~al.}, ``{System size
  dependence of cluster properties from two-particle angular correlations in
  Cu+Cu and Au+Au collisions at $\sqrt{s_{\rm NN}} = 200$ GeV}'',
  \href{http://dx.doi.org/10.1103/PhysRevC.81.024904}{{\em Phys. Rev. C}
  {\bfseries 81} (2010) 024904},
  \href{http://arxiv.org/abs/0812.1172}{{\ttfamily arXiv:0812.1172 [nucl-ex]}}.

\bibitem{STAR:2004wfp}
{\bfseries STAR} Collaboration, J.~Adams {\em et~al.}, ``{Minijet deformation
  and charge-independent angular correlations on momentum subspace ($\eta$,
  $\varphi$) in Au--Au collisions at $\sqrt{s_{\rm NN}} = 130$ \GeV}'',
  \href{http://dx.doi.org/10.1103/PhysRevC.73.064907}{{\em Phys. Rev. C}
  {\bfseries 73} (2006) 064907},
  \href{http://arxiv.org/abs/nucl-ex/0411003}{{\ttfamily
  arXiv:nucl-ex/0411003}}.

\bibitem{PHOBOS:2009sau}
{\bfseries PHOBOS} Collaboration, B.~Alver {\em et~al.}, ``{High transverse
  momentum triggered correlations over a large pseudorapidity acceptance in
  Au+Au collisions at $\sqrt{s_{\rm NN}} = 200$ \GeV}'',
  \href{http://dx.doi.org/10.1103/PhysRevLett.104.062301}{{\em Phys. Rev.
  Lett.} {\bfseries 104} (2010) 062301},
  \href{http://arxiv.org/abs/0903.2811}{{\ttfamily arXiv:0903.2811 [nucl-ex]}}.

\bibitem{STAR:2009ngv}
{\bfseries STAR} Collaboration, B.~I. Abelev {\em et~al.}, ``{Long range
  rapidity correlations and jet production in high energy nuclear
  collisions}'', \href{http://dx.doi.org/10.1103/PhysRevC.80.064912}{{\em Phys.
  Rev. C} {\bfseries 80} (2009) 064912},
  \href{http://arxiv.org/abs/0909.0191}{{\ttfamily arXiv:0909.0191 [nucl-ex]}}.

\bibitem{STAR:2011ryj}
{\bfseries STAR} Collaboration, G.~Agakishiev {\em et~al.}, ``{Anomalous
  centrality evolution of two-particle angular correlations from Au-Au
  collisions at $\sqrt{s_{\rm NN}} = 62$ and 200 GeV}'',
  \href{http://dx.doi.org/10.1103/PhysRevC.86.064902}{{\em Phys. Rev. C}
  {\bfseries 86} (2012) 064902},
  \href{http://arxiv.org/abs/1109.4380}{{\ttfamily arXiv:1109.4380 [nucl-ex]}}.

\bibitem{CMS:2011cqy}
{\bfseries CMS} Collaboration, S.~Chatrchyan {\em et~al.}, ``{Long-range and
  short-range dihadron angular correlations in central PbPb collisions at a
  nucleon-nucleon center of mass energy of 2.76 TeV}'',
  \href{http://dx.doi.org/10.1007/JHEP07(2011)076}{{\em JHEP} {\bfseries 07}
  (2011) 076}, \href{http://arxiv.org/abs/1105.2438}{{\ttfamily arXiv:1105.2438
  [nucl-ex]}}.

\bibitem{ALICE:2011svq}
{\bfseries ALICE} Collaboration, K.~Aamodt {\em et~al.}, ``{Harmonic
  decomposition of two-particle angular correlations in Pb-Pb collisions at
  $\sqrt{s_{\rm NN}}=2.76$ TeV}'',
  \href{http://dx.doi.org/10.1016/j.physletb.2012.01.060}{{\em Phys. Lett. B}
  {\bfseries 708} (2012) 249--264},
  \href{http://arxiv.org/abs/1109.2501}{{\ttfamily arXiv:1109.2501 [nucl-ex]}}.

\bibitem{CMS:2012xss}
{\bfseries CMS} Collaboration, S.~Chatrchyan {\em et~al.}, ``{Centrality
  dependence of dihadron correlations and azimuthal anisotropy harmonics in
  PbPb collisions at $\sqrt{s_{\rm NN}}=2.76$ TeV}'',
  \href{http://dx.doi.org/10.1140/epjc/s10052-012-2012-3}{{\em Eur. Phys. J. C}
  {\bfseries 72} (2012) 2012}, \href{http://arxiv.org/abs/1201.3158}{{\ttfamily
  arXiv:1201.3158 [nucl-ex]}}.

\bibitem{ATLAS:2012at}
{\bfseries ATLAS} Collaboration, G.~Aad {\em et~al.}, ``{Measurement of the
  azimuthal anisotropy for charged particle production in $\sqrt{s_{\rm NN}}=
  2.76$ TeV lead-lead collisions with the ATLAS detector}'',
  \href{http://dx.doi.org/10.1103/PhysRevC.86.014907}{{\em Phys. Rev. C}
  {\bfseries 86} (2012) 014907},
  \href{http://arxiv.org/abs/1203.3087}{{\ttfamily arXiv:1203.3087 [hep-ex]}}.

\bibitem{ALICE:2011ab}
{\bfseries ALICE} Collaboration, K.~Aamodt {\em et~al.}, ``{Higher harmonic
  anisotropic flow measurements of charged particles in Pb--Pb collisions at
  $\sqrt{s_{\rm NN}}=2.76$ TeV}'',
  \href{http://dx.doi.org/10.1103/PhysRevLett.107.032301}{{\em Phys. Rev.
  Lett.} {\bfseries 107} (2011) 032301},
  \href{http://arxiv.org/abs/1105.3865}{{\ttfamily arXiv:1105.3865 [nucl-ex]}}.

\bibitem{PhysRevD.46.229}
J.-Y. Ollitrault, ``Anisotropy as a signature of transverse collective flow'',
  \href{http://dx.doi.org/10.1103/PhysRevD.46.229}{{\em Phys. Rev. D}
  {\bfseries 46} (1992) 229--245}.

\bibitem{PhysRevC.81.054905}
B.~Alver and G.~Roland, ``Collision-geometry fluctuations and triangular flow
  in heavy-ion collisions'',
  \href{http://dx.doi.org/10.1103/PhysRevC.81.054905}{{\em Phys. Rev. C}
  {\bfseries 81} (2010) 054905}.

\bibitem{Denicol:2015nhu}
G.~Denicol, A.~Monnai, and B.~Schenke, ``{Moving forward to constrain the shear
  viscosity of QCD matter}'',
  \href{http://dx.doi.org/10.1103/PhysRevLett.116.212301}{{\em Phys. Rev.
  Lett.} {\bfseries 116} (2016) 212301},
  \href{http://arxiv.org/abs/1512.01538}{{\ttfamily arXiv:1512.01538
  [nucl-th]}}.

\bibitem{Nonaka:2017xsp}
C.~Nonaka and K.~Okamoto, ``{Phenomenological Analysis of High-Energy Heavy-Ion
  Collisions}'', \href{http://dx.doi.org/10.22323/1.294.0014}{{\em PoS}
  {\bfseries KMI2017} (2017) 014}.

\bibitem{Moreland:2018gsh}
J.~S. Moreland, J.~E. Bernhard, and S.~A. Bass, ``{Bayesian calibration of a
  hybrid nuclear collision model using p-Pb and Pb--Pb data at energies
  available at the CERN Large Hadron Collider}'',
  \href{http://dx.doi.org/10.1103/PhysRevC.101.024911}{{\em Phys. Rev. C}
  {\bfseries 101} (2020) 024911},
  \href{http://arxiv.org/abs/1808.02106}{{\ttfamily arXiv:1808.02106
  [nucl-th]}}.

\bibitem{PHOBOS:2004nvy}
{\bfseries PHOBOS} Collaboration, B.~B. Back {\em et~al.}, ``{Energy dependence
  of elliptic flow over a large pseudorapidity range in Au+Au collisions at
  RHIC}'', \href{http://dx.doi.org/10.1103/PhysRevLett.94.122303}{{\em Phys.
  Rev. Lett.} {\bfseries 94} (2005) 122303},
  \href{http://arxiv.org/abs/nucl-ex/0406021}{{\ttfamily
  arXiv:nucl-ex/0406021}}.

\bibitem{ALICE:2016tlx}
{\bfseries ALICE} Collaboration, J.~Adam {\em et~al.}, ``{Pseudorapidity
  dependence of the anisotropic flow of charged particles in Pb--Pb collisions
  at $\sqrt{s_{\rm NN}}=2.76$ TeV}'',
  \href{http://dx.doi.org/10.1016/j.physletb.2016.07.017}{{\em Phys. Lett. B}
  {\bfseries 762} (2016) 376--388},
  \href{http://arxiv.org/abs/1605.02035}{{\ttfamily arXiv:1605.02035
  [nucl-ex]}}.

\bibitem{PHENIX:2013ktj}
{\bfseries PHENIX} Collaboration, A.~Adare {\em et~al.}, ``{Quadrupole
  Anisotropy in Dihadron Azimuthal Correlations in Central $d+$Au Collisions at
  $\sqrt{s_{\rm NN}}=200$ GeV}'',
  \href{http://dx.doi.org/10.1103/PhysRevLett.111.212301}{{\em Phys. Rev.
  Lett.} {\bfseries 111} (2013) 212301},
  \href{http://arxiv.org/abs/1303.1794}{{\ttfamily arXiv:1303.1794 [nucl-ex]}}.

\bibitem{PHENIX:2016cfs}
{\bfseries PHENIX} Collaboration, C.~Aidala {\em et~al.}, ``{Measurement of
  long-range angular correlations and azimuthal anisotropies in
  high-multiplicity $p+$Au collisions at $\sqrt{s_{\rm NN}}=200$ GeV}'',
  \href{http://dx.doi.org/10.1103/PhysRevC.95.034910}{{\em Phys. Rev. C}
  {\bfseries 95} (2017) 034910},
  \href{http://arxiv.org/abs/1609.02894}{{\ttfamily arXiv:1609.02894
  [nucl-ex]}}.

\bibitem{PHENIX:2021ubk}
{\bfseries PHENIX} Collaboration, U.~A. Acharya {\em et~al.}, ``{Kinematic
  dependence of azimuthal anisotropies in $p$+Au, $d$+Au, and $^{3}$He+Au at
  $\sqrt{s_{\rm NN}}=200$ GeV}'',
  \href{http://dx.doi.org/10.1103/PhysRevC.105.024901}{{\em Phys. Rev. C}
  {\bfseries 105} (2022) 024901},
  \href{http://arxiv.org/abs/2107.06634}{{\ttfamily arXiv:2107.06634
  [hep-ex]}}.

\bibitem{STAR:2014qsy}
{\bfseries STAR} Collaboration, L.~Adamczyk {\em et~al.}, ``{Effect of event
  selection on jetlike correlation measurement in $d$+Au collisions at
  $\sqrt{s_{\rm{NN}}}=200$ GeV}'',
  \href{http://dx.doi.org/10.1016/j.physletb.2015.02.068}{{\em Phys. Lett. B}
  {\bfseries 743} (2015) 333--339},
  \href{http://arxiv.org/abs/1412.8437}{{\ttfamily arXiv:1412.8437 [nucl-ex]}}.

\bibitem{STAR:2022pfn}
{\bfseries STAR} Collaboration, M.~I. Abdulhamid {\em et~al.}, ``{Measurements
  of the Elliptic and Triangular Azimuthal Anisotropies in Central He3+Au, d+Au
  and p+Au Collisions at sNN=200\,\,GeV}'',
  \href{http://dx.doi.org/10.1103/PhysRevLett.130.242301}{{\em Phys. Rev.
  Lett.} {\bfseries 130} (2023) 242301},
  \href{http://arxiv.org/abs/2210.11352}{{\ttfamily arXiv:2210.11352
  [nucl-ex]}}.

\bibitem{CMS:2010ifv}
{\bfseries CMS} Collaboration, V.~Khachatryan {\em et~al.}, ``{Observation of
  Long-Range Near-Side Angular Correlations in Proton-Proton Collisions at the
  LHC}'', \href{http://dx.doi.org/10.1007/JHEP09(2010)091}{{\em JHEP}
  {\bfseries 09} (2010) 091}, \href{http://arxiv.org/abs/1009.4122}{{\ttfamily
  arXiv:1009.4122 [hep-ex]}}.

\bibitem{CMS:2012qk}
{\bfseries CMS} Collaboration, S.~Chatrchyan {\em et~al.}, ``{Observation of
  Long-Range Near-Side Angular Correlations in Proton-Lead Collisions at the
  LHC}'', \href{http://dx.doi.org/10.1016/j.physletb.2012.11.025}{{\em Phys.
  Lett. B} {\bfseries 718} (2013) 795--814},
  \href{http://arxiv.org/abs/1210.5482}{{\ttfamily arXiv:1210.5482 [nucl-ex]}}.

\bibitem{CMS:2015fgy}
{\bfseries CMS} Collaboration, V.~Khachatryan {\em et~al.}, ``{Measurement of
  long-range near-side two-particle angular correlations in pp collisions at
  $\sqrt s =13$ TeV}'',
  \href{http://dx.doi.org/10.1103/PhysRevLett.116.172302}{{\em Phys. Rev.
  Lett.} {\bfseries 116} (2016) 172302},
  \href{http://arxiv.org/abs/1510.03068}{{\ttfamily arXiv:1510.03068
  [nucl-ex]}}.

\bibitem{CMS:2016fnw}
{\bfseries CMS} Collaboration, V.~Khachatryan {\em et~al.}, ``{Evidence for
  collectivity in pp collisions at the LHC}'',
  \href{http://dx.doi.org/10.1016/j.physletb.2016.12.009}{{\em Phys. Lett. B}
  {\bfseries 765} (2017) 193--220},
  \href{http://arxiv.org/abs/1606.06198}{{\ttfamily arXiv:1606.06198
  [nucl-ex]}}.

\bibitem{ALICE:2012eyl}
{\bfseries ALICE} Collaboration, B.~Abelev {\em et~al.}, ``{Long-range angular
  correlations on the near and away side in $p$-Pb collisions at $\sqrt{s_{\rm
  NN}}=5.02$ TeV}'',
  \href{http://dx.doi.org/10.1016/j.physletb.2013.01.012}{{\em Phys. Lett. B}
  {\bfseries 719} (2013) 29--41},
  \href{http://arxiv.org/abs/1212.2001}{{\ttfamily arXiv:1212.2001 [nucl-ex]}}.

\bibitem{ALICE:2013snk}
{\bfseries ALICE} Collaboration, B.~Abelev {\em et~al.}, ``{Long-range angular
  correlations of $\rm \pi$, K and p in p-Pb collisions at $\sqrt{s_{\rm NN}} =
  5.02$ TeV}'', \href{http://dx.doi.org/10.1016/j.physletb.2013.08.024}{{\em
  Phys. Lett. B} {\bfseries 726} (2013) 164--177},
  \href{http://arxiv.org/abs/1307.3237}{{\ttfamily arXiv:1307.3237 [nucl-ex]}}.

\bibitem{ATLAS:2012cix}
{\bfseries ATLAS} Collaboration, G.~Aad {\em et~al.}, ``{Observation of
  Associated Near-Side and Away-Side Long-Range Correlations in $\sqrt{s_{\rm
  NN}}=5.02$ TeV Proton-Lead Collisions with the ATLAS Detector}'',
  \href{http://dx.doi.org/10.1103/PhysRevLett.110.182302}{{\em Phys. Rev.
  Lett.} {\bfseries 110} (2013) 182302},
  \href{http://arxiv.org/abs/1212.5198}{{\ttfamily arXiv:1212.5198 [hep-ex]}}.

\bibitem{ATLAS:2014qaj}
{\bfseries ATLAS} Collaboration, G.~Aad {\em et~al.}, ``{Measurement of
  long-range pseudorapidity correlations and azimuthal harmonics in
  $\sqrt{s_{\rm NN}}=5.02$ TeV proton-lead collisions with the ATLAS
  detector}'', \href{http://dx.doi.org/10.1103/PhysRevC.90.044906}{{\em Phys.
  Rev. C} {\bfseries 90} (2014) 044906},
  \href{http://arxiv.org/abs/1409.1792}{{\ttfamily arXiv:1409.1792 [hep-ex]}}.

\bibitem{CMS:2018loe}
{\bfseries CMS} Collaboration, A.~M. Sirunyan {\em et~al.}, ``{Elliptic flow of
  charm and strange hadrons in high-multiplicity pPb collisions at
  $\sqrt{s_{\mathrm{NN}}}= 8.16$ TeV}'',
  \href{http://dx.doi.org/10.1103/PhysRevLett.121.082301}{{\em Phys. Rev.
  Lett.} {\bfseries 121} (2018) 082301},
  \href{http://arxiv.org/abs/1804.09767}{{\ttfamily arXiv:1804.09767
  [hep-ex]}}.

\bibitem{Bozek:2013uha}
P.~Bozek and W.~Broniowski, ``{Collective dynamics in high-energy
  proton-nucleus collisions}'',
  \href{http://dx.doi.org/10.1103/PhysRevC.88.014903}{{\em Phys. Rev. C}
  {\bfseries 88} (2013) 014903},
  \href{http://arxiv.org/abs/1304.3044}{{\ttfamily arXiv:1304.3044 [nucl-th]}}.

\bibitem{Weller:2017tsr}
R.~D. Weller and P.~Romatschke, ``{One fluid to rule them all: viscous
  hydrodynamic description of event-by-event central p+p, p+Pb and Pb+Pb
  collisions at $\sqrt{s}=5.02$ TeV}'',
  \href{http://dx.doi.org/10.1016/j.physletb.2017.09.077}{{\em Phys. Lett. B}
  {\bfseries 774} (2017) 351--356},
  \href{http://arxiv.org/abs/1701.07145}{{\ttfamily arXiv:1701.07145
  [nucl-th]}}.

\bibitem{Albacete:2013ei}
J.~L. Albacete {\em et~al.}, ``{Predictions for $p+$Pb Collisions at
  $\sqrt{s_{\rm NN}} = 5$ TeV}'',
  \href{http://dx.doi.org/10.1142/S0218301313300075}{{\em Int. J. Mod. Phys. E}
  {\bfseries 22} (2013) 1330007},
  \href{http://arxiv.org/abs/1301.3395}{{\ttfamily arXiv:1301.3395 [hep-ph]}}.

\bibitem{Albacete:2016veq}
J.~L. Albacete {\em et~al.}, ``{Predictions for $p+$Pb Collisions at
  $\sqrt{s_{NN}} = 5$ TeV: Comparison with Data}'',
  \href{http://dx.doi.org/10.1142/S0218301316300058}{{\em Int. J. Mod. Phys. E}
  {\bfseries 25} (2016) 1630005},
  \href{http://arxiv.org/abs/1605.09479}{{\ttfamily arXiv:1605.09479
  [hep-ph]}}.

\bibitem{Zhou:2015iba}
Y.~Zhou, X.~Zhu, P.~Li, and H.~Song, ``{Investigation of possible hadronic flow
  in $\sqrt{s_{NN}} = 5.02$ TeV $p-Pb$ collisions}'',
  \href{http://dx.doi.org/10.1103/PhysRevC.91.064908}{{\em Phys. Rev. C}
  {\bfseries 91} (2015) 064908},
  \href{http://arxiv.org/abs/1503.06986}{{\ttfamily arXiv:1503.06986
  [nucl-th]}}.

\bibitem{Zhao:2017rgg}
W.~Zhao, Y.~Zhou, H.~Xu, W.~Deng, and H.~Song, ``{Hydrodynamic collectivity in
  proton\textendash{}proton collisions at 13 TeV}'',
  \href{http://dx.doi.org/10.1016/j.physletb.2018.03.022}{{\em Phys. Lett. B}
  {\bfseries 780} (2018) 495--500},
  \href{http://arxiv.org/abs/1801.00271}{{\ttfamily arXiv:1801.00271
  [nucl-th]}}.

\bibitem{Schenke:2019pmk}
B.~Schenke, C.~Shen, and P.~Tribedy, ``{Hybrid Color Glass Condensate and
  hydrodynamic description of the Relativistic Heavy Ion Collider small system
  scan}'', \href{http://dx.doi.org/10.1016/j.physletb.2020.135322}{{\em Phys.
  Lett. B} {\bfseries 803} (2020) 135322},
  \href{http://arxiv.org/abs/1908.06212}{{\ttfamily arXiv:1908.06212
  [nucl-th]}}.

\bibitem{Kanakubo:2021qcw}
Y.~Kanakubo, Y.~Tachibana, and T.~Hirano, ``{Interplay between core and corona
  components in high-energy nuclear collisions}'',
  \href{http://dx.doi.org/10.1103/PhysRevC.105.024905}{{\em Phys. Rev. C}
  {\bfseries 105} (2022) 024905},
  \href{http://arxiv.org/abs/2108.07943}{{\ttfamily arXiv:2108.07943
  [nucl-th]}}.

\bibitem{Nie:2019swk}
M.~Nie, L.~Yi, G.~Ma, and J.~Jia, ``{Influence of initial-state momentum
  anisotropy on the final-state collectivity in small collision systems}'',
  \href{http://dx.doi.org/10.1103/PhysRevC.100.064905}{{\em Phys. Rev. C}
  {\bfseries 100} (2019) 064905},
  \href{http://arxiv.org/abs/1906.01422}{{\ttfamily arXiv:1906.01422
  [nucl-th]}}.

\bibitem{ALICE:2022imr}
{\bfseries ALICE} Collaboration, S.~Acharya {\em et~al.}, ``{System-size
  dependence of the charged-particle pseudorapidity density at $\sqrt{s_{\rm
  NN}} = 5.02$ TeV for pp, p-Pb, and Pb-Pb collisions}'',
  \href{http://dx.doi.org/10.1016/j.physletb.2023.137730}{{\em Phys. Lett. B}
  {\bfseries 845} (2023) 137730},
  \href{http://arxiv.org/abs/2204.10210}{{\ttfamily arXiv:2204.10210
  [nucl-ex]}}.

\bibitem{CMS:2016est}
{\bfseries CMS} Collaboration, V.~Khachatryan {\em et~al.}, ``{Pseudorapidity
  dependence of long-range two-particle correlations in $p$Pb collisions at
  $\sqrt {s_{\rm NN}}= 5.02$ TeV}'',
  \href{http://dx.doi.org/10.1103/PhysRevC.96.014915}{{\em Phys. Rev. C}
  {\bfseries 96} (2017) 014915},
  \href{http://arxiv.org/abs/1604.05347}{{\ttfamily arXiv:1604.05347
  [nucl-ex]}}.

\bibitem{CMS:2017xnj}
{\bfseries CMS} Collaboration, A.~M. Sirunyan {\em et~al.}, ``{Pseudorapidity
  and transverse momentum dependence of flow harmonics in pPb and PbPb
  collisions}'', \href{http://dx.doi.org/10.1103/PhysRevC.98.044902}{{\em Phys.
  Rev. C} {\bfseries 98} (2018) 044902},
  \href{http://arxiv.org/abs/1710.07864}{{\ttfamily arXiv:1710.07864
  [nucl-ex]}}.

\bibitem{PHENIX:2018hho}
{\bfseries PHENIX} Collaboration, A.~Adare {\em et~al.}, ``{Pseudorapidity
  Dependence of Particle Production and Elliptic Flow in Asymmetric Nuclear
  Collisions of $p+$Al, $p+$Au, $d+$Au, and $^{3}$He$+$Au at $\sqrt{s_{\rm
  NN}}=200$ GeV}'',
  \href{http://dx.doi.org/10.1103/PhysRevLett.121.222301}{{\em Phys. Rev.
  Lett.} {\bfseries 121} (2018) 222301},
  \href{http://arxiv.org/abs/1807.11928}{{\ttfamily arXiv:1807.11928
  [nucl-ex]}}.

\bibitem{PHENIX:2017nae}
{\bfseries PHENIX} Collaboration, C.~Aidala {\em et~al.}, ``{Measurements of
  azimuthal anisotropy and charged-particle multiplicity in $d+$Au collisions
  at $\sqrt{s_{\rm NN}}=200$, 62.4, 39, and 19.6 GeV}'',
  \href{http://dx.doi.org/10.1103/PhysRevC.96.064905}{{\em Phys. Rev. C}
  {\bfseries 96} (2017) 064905},
  \href{http://arxiv.org/abs/1708.06983}{{\ttfamily arXiv:1708.06983
  [nucl-ex]}}.

\bibitem{Bozek:2014cya}
P.~Bozek and W.~Broniowski, ``{Collective flow in ultrarelativistic $^3$He-Au
  collisions}'', \href{http://dx.doi.org/10.1016/j.physletb.2014.11.006}{{\em
  Phys. Lett. B} {\bfseries 739} (2014) 308--312},
  \href{http://arxiv.org/abs/1409.2160}{{\ttfamily arXiv:1409.2160 [nucl-th]}}.

\bibitem{ALICE:2008ngc}
{\bfseries ALICE} Collaboration, K.~Aamodt {\em et~al.}, ``{The ALICE
  experiment at the CERN LHC}'',
  \href{http://dx.doi.org/10.1088/1748-0221/3/08/S08002}{{\em JINST} {\bfseries
  3} (2008) S08002}.

\bibitem{ALICE:2014sbx}
{\bfseries ALICE} Collaboration, B.~B. Abelev {\em et~al.}, ``{Performance of
  the ALICE Experiment at the CERN LHC}'',
  \href{http://dx.doi.org/10.1142/S0217751X14300440}{{\em Int. J. Mod. Phys. A}
  {\bfseries 29} (2014) 1430044},
  \href{http://arxiv.org/abs/1402.4476}{{\ttfamily arXiv:1402.4476 [nucl-ex]}}.

\bibitem{ALICE:2012xs}
{\bfseries ALICE} Collaboration, B.~Abelev {\em et~al.}, ``{Pseudorapidity
  density of charged particles in $p$ + Pb collisions at $\sqrt{s_{\rm
  NN}}=5.02$ TeV}'',
  \href{http://dx.doi.org/10.1103/PhysRevLett.110.032301}{{\em Phys. Rev.
  Lett.} {\bfseries 110} (2013) 032301},
  \href{http://arxiv.org/abs/1210.3615}{{\ttfamily arXiv:1210.3615 [nucl-ex]}}.

\bibitem{ALICE:2012aqc}
{\bfseries ALICE} Collaboration, B.~Abelev {\em et~al.}, ``{Centrality
  Dependence of Charged Particle Production at Large Transverse Momentum in
  Pb--Pb Collisions at $\sqrt{s_{\rm{NN}}} = 2.76$ TeV}'',
  \href{http://dx.doi.org/10.1016/j.physletb.2013.01.051}{{\em Phys. Lett. B}
  {\bfseries 720} (2013) 52--62},
  \href{http://arxiv.org/abs/1208.2711}{{\ttfamily arXiv:1208.2711 [hep-ex]}}.

\bibitem{Roesler:2000he}
S.~Roesler, R.~Engel, and J.~Ranft,
  \href{http://dx.doi.org/10.1007/978-3-642-18211-2_166}{``{The Monte Carlo
  event generator DPMJET-III}'',} in {\em {International Conference on Advanced
  Monte Carlo for Radiation Physics, Particle Transport Simulation and
  Applications (MC 2000)}}, pp.~1033--1038.
\newblock 12, 2000.
\newblock \href{http://arxiv.org/abs/hep-ph/0012252}{{\ttfamily
  arXiv:hep-ph/0012252}}.

\bibitem{Brun:1082634}
R.~Brun, F.~Bruyant, F.~Carminati, S.~Giani, M.~Maire, A.~McPherson,
  G.~Patrick, and L.~Urban,
  \href{http://dx.doi.org/10.17181/CERN.MUHF.DMJ1}{{\em {GEANT: Detector
  Description and Simulation Tool; Oct 1994}}}.
\newblock CERN Program Library. CERN, Geneva, 1993.
\newblock \url{https://cds.cern.ch/record/1082634}.
\newblock Long Writeup W5013.

\bibitem{ATLAS:2015hzw}
{\bfseries ATLAS} Collaboration, G.~Aad {\em et~al.}, ``{Observation of
  Long-Range Elliptic Azimuthal Anisotropies in $\sqrt{s}=$13 and 2.76 TeV $pp$
  Collisions with the ATLAS Detector}'',
  \href{http://dx.doi.org/10.1103/PhysRevLett.116.172301}{{\em Phys. Rev.
  Lett.} {\bfseries 116} (2016) 172301},
  \href{http://arxiv.org/abs/1509.04776}{{\ttfamily arXiv:1509.04776
  [hep-ex]}}.

\bibitem{ATLAS:2016yzd}
{\bfseries ATLAS} Collaboration, M.~Aaboud {\em et~al.}, ``{Measurements of
  long-range azimuthal anisotropies and associated Fourier coefficients for
  $pp$ collisions at $\sqrt{s}=5.02$ and $13$ TeV and $p$+Pb collisions at
  $\sqrt{s_{\mathrm{NN}}}=5.02$ TeV with the ATLAS detector}'',
  \href{http://dx.doi.org/10.1103/PhysRevC.96.024908}{{\em Phys. Rev. C}
  {\bfseries 96} (2017) 024908},
  \href{http://arxiv.org/abs/1609.06213}{{\ttfamily arXiv:1609.06213
  [nucl-ex]}}.

\bibitem{PHENIX:2022nht}
{\bfseries PHENIX} Collaboration, N.~J. Abdulameer {\em et~al.},
  ``{Measurements of second-harmonic Fourier coefficients from azimuthal
  anisotropies in $p+p$, $p$+Au, $d$+Au, and $^3$He+Au collisions at
  $\sqrt{s_{\rm NN}}=200$GeV}'',
  \href{http://dx.doi.org/10.1103/PhysRevC.107.024907}{{\em Phys. Rev. C}
  {\bfseries 107} (2023) 024907},
  \href{http://arxiv.org/abs/2203.09894}{{\ttfamily arXiv:2203.09894
  [nucl-ex]}}.

\bibitem{ATLAS:2017rij}
{\bfseries ATLAS} Collaboration, M.~Aaboud {\em et~al.}, ``{Measurement of
  longitudinal flow decorrelations in Pb+Pb collisions at $\sqrt{s_{\text
  {NN}}}=2.76$ and 5.02 TeV with the ATLAS detector}'',
  \href{http://dx.doi.org/10.1140/epjc/s10052-018-5605-7}{{\em Eur. Phys. J. C}
  {\bfseries 78} (2018) 142}, \href{http://arxiv.org/abs/1709.02301}{{\ttfamily
  arXiv:1709.02301 [nucl-ex]}}.

\bibitem{ATLAS:2020sgl}
{\bfseries ATLAS} Collaboration, G.~Aad {\em et~al.}, ``{Longitudinal Flow
  Decorrelations in Xe+Xe Collisions at $\sqrt{s_{\mathrm{NN}}}=5.44$ TeV with
  the ATLAS Detector}'',
  \href{http://dx.doi.org/10.1103/PhysRevLett.126.122301}{{\em Phys. Rev.
  Lett.} {\bfseries 126} (2021) 122301},
  \href{http://arxiv.org/abs/2001.04201}{{\ttfamily arXiv:2001.04201
  [nucl-ex]}}.

\bibitem{CMS:2015xmx}
{\bfseries CMS} Collaboration, V.~Khachatryan {\em et~al.}, ``{Evidence for
  transverse momentum and pseudorapidity dependent event plane fluctuations in
  PbPb and pPb collisions}'',
  \href{http://dx.doi.org/10.1103/PhysRevC.92.034911}{{\em Phys. Rev. C}
  {\bfseries 92} (2015) 034911},
  \href{http://arxiv.org/abs/1503.01692}{{\ttfamily arXiv:1503.01692
  [nucl-ex]}}.

\bibitem{Sakai:2021rug}
A.~Sakai, K.~Murase, and T.~Hirano, ``{Effects of hydrodynamic and initial
  longitudinal fluctuations on rapidity decorrelation of collective flow}'',
  \href{http://dx.doi.org/10.1016/j.physletb.2022.137053}{{\em Phys. Lett. B}
  {\bfseries 829} (2022) 137053},
  \href{http://arxiv.org/abs/2111.08963}{{\ttfamily arXiv:2111.08963
  [nucl-th]}}.

\bibitem{Bozek:2017qir}
P.~Bozek and W.~Broniowski, ``{Longitudinal decorrelation measures of flow
  magnitude and event-plane angles in ultrarelativistic nuclear collisions}'',
  \href{http://dx.doi.org/10.1103/PhysRevC.97.034913}{{\em Phys. Rev. C}
  {\bfseries 97} (2018) 034913},
  \href{http://arxiv.org/abs/1711.03325}{{\ttfamily arXiv:1711.03325
  [nucl-th]}}.

\bibitem{Bozek:2015bna}
P.~Bozek and W.~Broniowski, ``{The torque effect and fluctuations of entropy
  deposition in rapidity in ultra-relativistic nuclear collisions}'',
  \href{http://dx.doi.org/10.1016/j.physletb.2015.11.054}{{\em Phys. Lett. B}
  {\bfseries 752} (2016) 206--211},
  \href{http://arxiv.org/abs/1506.02817}{{\ttfamily arXiv:1506.02817
  [nucl-th]}}.

\bibitem{Lin:2004en}
Z.-W. Lin, C.~M. Ko, B.-A. Li, B.~Zhang, and S.~Pal, ``{A Multi-phase transport
  model for relativistic heavy ion collisions}'',
  \href{http://dx.doi.org/10.1103/PhysRevC.72.064901}{{\em Phys. Rev. C}
  {\bfseries 72} (2005) 064901},
  \href{http://arxiv.org/abs/nucl-th/0411110}{{\ttfamily
  arXiv:nucl-th/0411110}}.

\bibitem{Pierog:2013ria}
T.~Pierog, I.~Karpenko, J.~M. Katzy, E.~Yatsenko, and K.~Werner, ``{EPOS LHC:
  Test of collective hadronization with data measured at the CERN Large Hadron
  Collider}'', \href{http://dx.doi.org/10.1103/PhysRevC.92.034906}{{\em Phys.
  Rev. C} {\bfseries 92} (2015) 034906},
  \href{http://arxiv.org/abs/1306.0121}{{\ttfamily arXiv:1306.0121 [hep-ph]}}.

\bibitem{Barlow:2002yb}
R.~Barlow, ``{Systematic errors: Facts and fictions}'', in {\em {Conference on
  Advanced Statistical Techniques in Particle Physics}}, pp.~134--144.
\newblock 7, 2002.
\newblock \href{http://arxiv.org/abs/hep-ex/0207026}{{\ttfamily
  arXiv:hep-ex/0207026}}.

\bibitem{ALICE:2015bpk}
{\bfseries ALICE} Collaboration, J.~Adam {\em et~al.}, ``{Centrality evolution
  of the charged\textendash{}particle pseudorapidity density over a broad
  pseudorapidity range in Pb\textendash{}Pb collisions at $\sqrt{s_{\rm
  NN}}=2.76$ TeV}'',
  \href{http://dx.doi.org/10.1016/j.physletb.2015.12.082}{{\em Phys. Lett. B}
  {\bfseries 754} (2016) 373--385},
  \href{http://arxiv.org/abs/1509.07299}{{\ttfamily arXiv:1509.07299
  [nucl-ex]}}.

\bibitem{Zhao:2022ayk}
W.~Zhao, C.~Shen, and B.~Schenke, ``{Collectivity in Ultraperipheral Pb+Pb
  Collisions at the Large Hadron Collider}'',
  \href{http://dx.doi.org/10.1103/PhysRevLett.129.252302}{{\em Phys. Rev.
  Lett.} {\bfseries 129} (2022) 252302},
  \href{http://arxiv.org/abs/2203.06094}{{\ttfamily arXiv:2203.06094
  [nucl-th]}}.

\bibitem{Shen:2022oyg}
C.~Shen and B.~Schenke, ``{Longitudinal dynamics and particle production in
  relativistic nuclear collisions}'',
  \href{http://dx.doi.org/10.1103/PhysRevC.105.064905}{{\em Phys. Rev. C}
  {\bfseries 105} (2022) 064905},
  \href{http://arxiv.org/abs/2203.04685}{{\ttfamily arXiv:2203.04685
  [nucl-th]}}.

\bibitem{Schenke:2022mjv}
B.~Schenke, S.~Schlichting, and P.~Singh, ``{Rapidity dependence of initial
  state geometry and momentum correlations in p+Pb collisions}'',
  \href{http://dx.doi.org/10.1103/PhysRevD.105.094023}{{\em Phys. Rev. D}
  {\bfseries 105} (2022) 094023},
  \href{http://arxiv.org/abs/2201.08864}{{\ttfamily arXiv:2201.08864
  [nucl-th]}}.

\bibitem{ATLAS:2018ngv}
{\bfseries ATLAS} Collaboration, M.~Aaboud {\em et~al.}, ``{Correlated
  long-range mixed-harmonic fluctuations measured in $pp$, $p$+Pb and
  low-multiplicity Pb+Pb collisions with the ATLAS detector}'',
  \href{http://dx.doi.org/10.1016/j.physletb.2018.11.065}{{\em Phys. Lett. B}
  {\bfseries 789} (2019) 444--471},
  \href{http://arxiv.org/abs/1807.02012}{{\ttfamily arXiv:1807.02012
  [nucl-ex]}}.

\end{thebibliography}\endgroup

\newpage
\appendix
\appendix
\section{Non-flow subtraction with the improved-template-fit method and peripheral subtraction procedure}
To examine the robustness of the observation, two different derived template-fit methods for non-flow suppression are employed: the zero-yield-at-minimum (ZYAM) and the improved-template-fit 
method~\cite{ATLAS:2018ngv, ATLAS:2015hzw}. 
Henceforth, the peripheral subtraction will denote the 
ZYAM template-fit method. 
The term of $FY^{\rm peri}(\Delta\varphi)$ in Eq.~\eqref{eqtemplate} is substituted by $FY^{\rm peri}(\Delta\varphi)-Y_{0}$, where $Y_{0}$ is the baseline of the correlation function in peripheral events, which is determined at $\Delta\varphi\sim0$.
The peripheral subtraction method assumes that the 2nd-order modulation is zero when the multiplicity is zero.
An improved-template-fit method was developed to estimate the non-flow effects in peripheral collisions more realistically. This method assumes that there are no flow components in the peripheral events. 
On the other hand, this procedure takes into account the residual flow components in peripheral events.
The flow harmonics in the second-most peripheral events are extracted by the template fit using the correlation function in the most peripheral events given by
\begin{equation}
\label{eqperitemp}
    Y^{\rm peri}(\Delta\varphi)=Y_{\rm non\mathchar`- flow}(\Delta\varphi)+G^{\rm peri}\lbrace1+2\sum_{n=2}^{3}v^{\rm peri}_{n,n}\cos{(n\Delta\varphi)}\rbrace,
\end{equation}
where $Y^{\rm peri}(\Delta\varphi)$ and $Y_{\rm non-flow}(\Delta\varphi)$ are the correlation functions in the second-most 
and the most peripheral events, respectively.
By substituting Eq. (A.1) with Eq. (3), it can be seen that
\begin{equation}
    Y(\Delta\varphi)=FY_{\rm non \mathchar`-
    flow}(\Delta\varphi)+(G^{\rm tmp}+FG^{\rm  peri})\lbrace1+2\sum_{n=2}^{3}\frac{G^{\rm tmp}v^{\rm tmp}_{n,n}+FG^{\rm peri}v^{\rm peri}_{n,n}}{G^{\rm tmp}+FG^{\rm peri}}\cos(n\Delta\varphi)\rbrace.
\end{equation}
Finally, the non-flow subtracted flow coefficient from the improved-template-fit method, $v^{\rm imp}_{n,n}$, can be obtained as
\begin{equation}
    v^{\rm imp}_{n,n}=v^{\rm tmp}_{n,n}-\frac{FG^{\rm peri}}{G^{\rm tmp}+FG^{\rm peri}}(v^{\rm tmp}_{n,n}-v^{\rm peri}_{n,n}).
\end{equation}

\begin{figure}[H]
    \begin{center}
    \includegraphics[width= 0.75\textwidth]{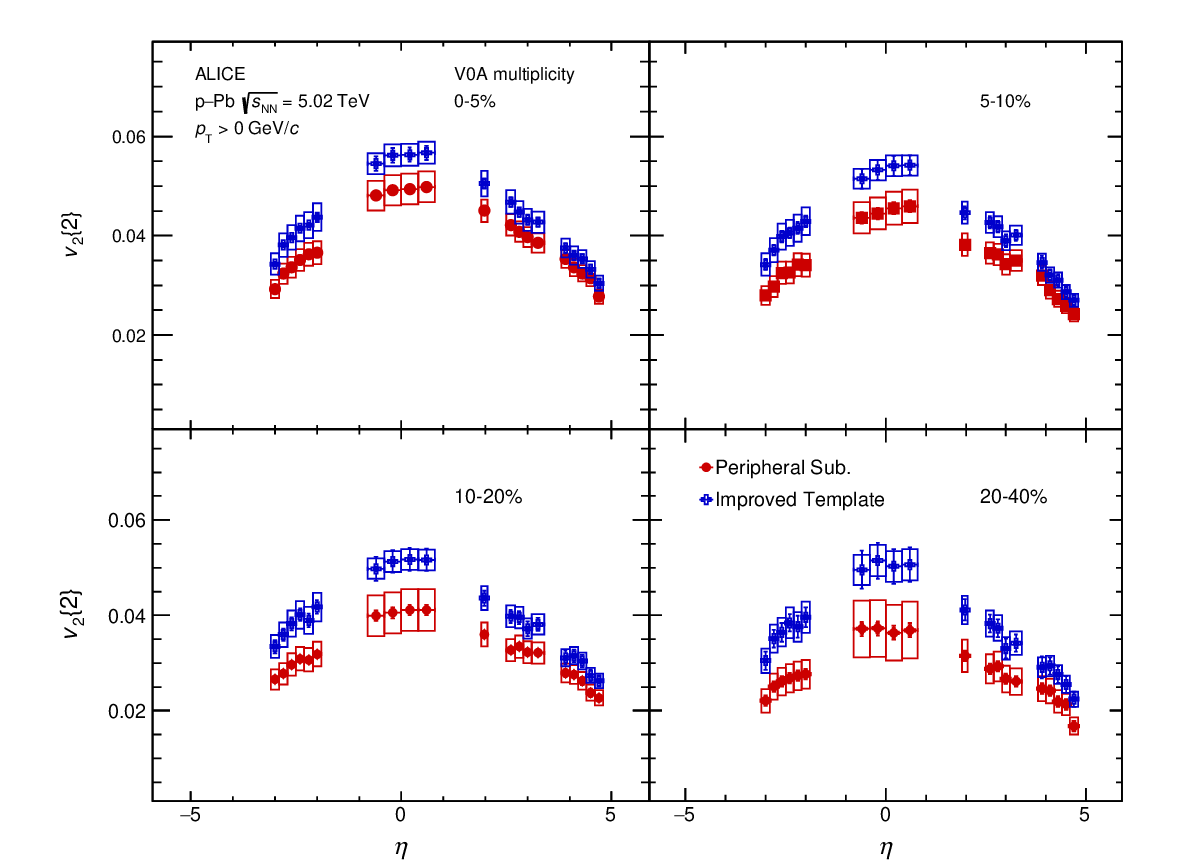}
    \end{center}
    \caption{$p_{\mathrm T}$-integrated $v_{2}\{2\}$ as a function of $\eta$ in various centrality classes using the improved-template-fit method, and the peripheral subtraction method.}
    \label{fig:compnonflow}
\end{figure}

\begin{figure}[H]
    \begin{center}
    \includegraphics[width= 0.48\textwidth]{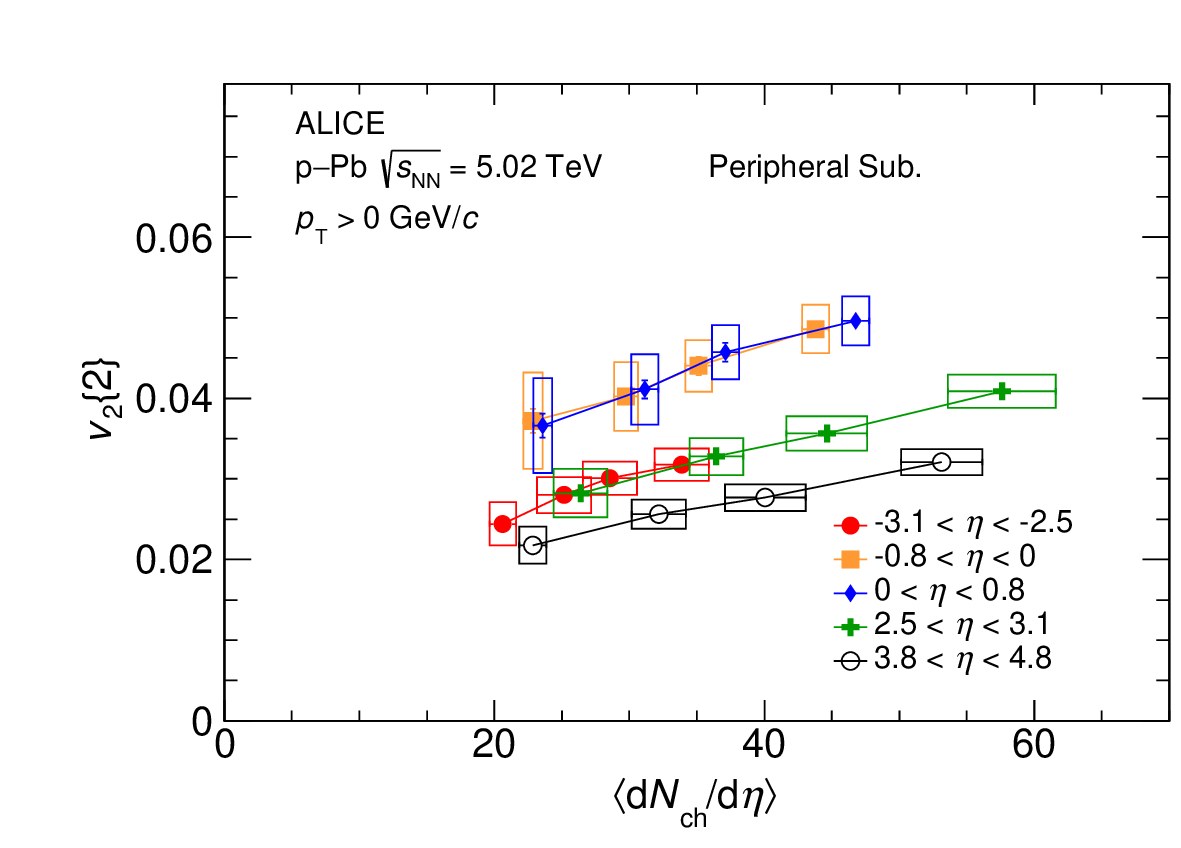}
     \includegraphics[width= 0.48\textwidth]{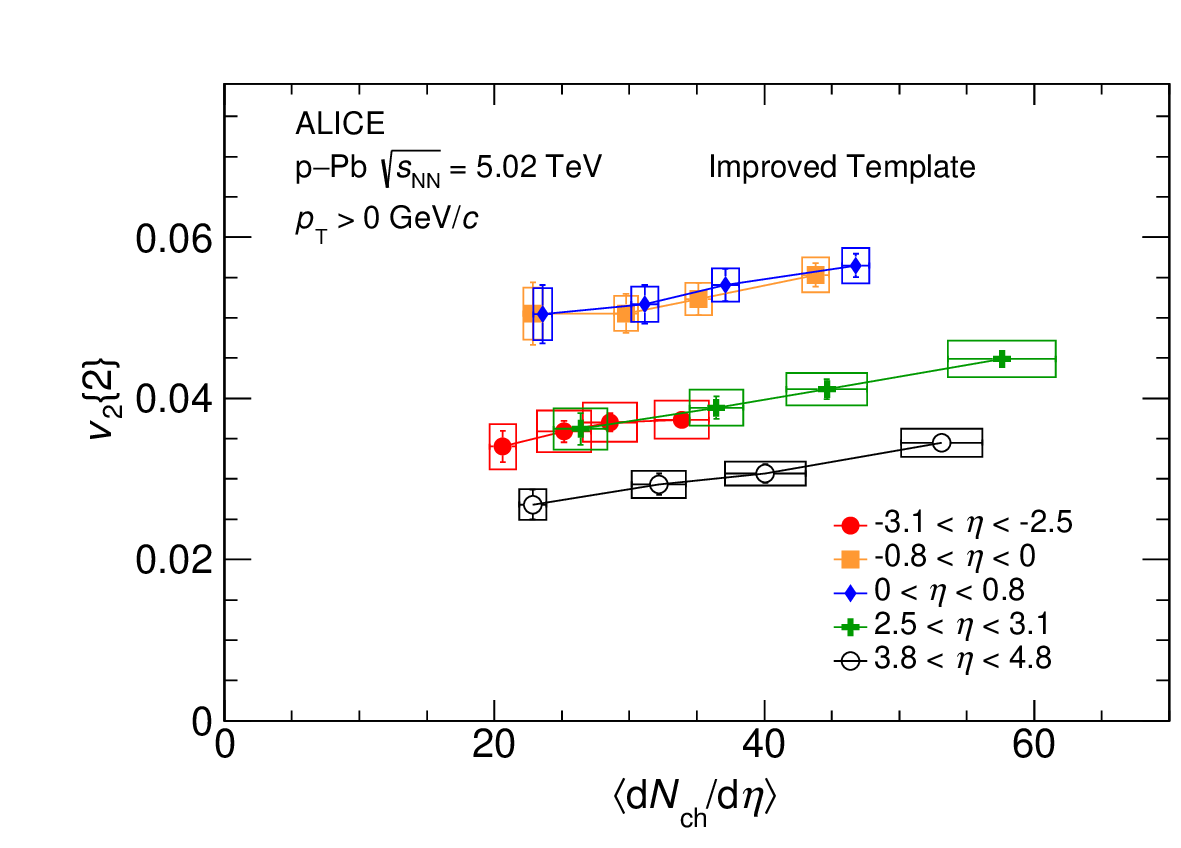}

    \end{center}
    \caption{\vtwo as a function of charged particle density for five different pseudorapidity regions with the peripheral subtraction (left) and the improved template method (right). }
    
    \label{fig:dndetaappendix}
\end{figure}

\begin{figure}[H]
    \begin{center}
    \includegraphics[width= 0.75\textwidth]{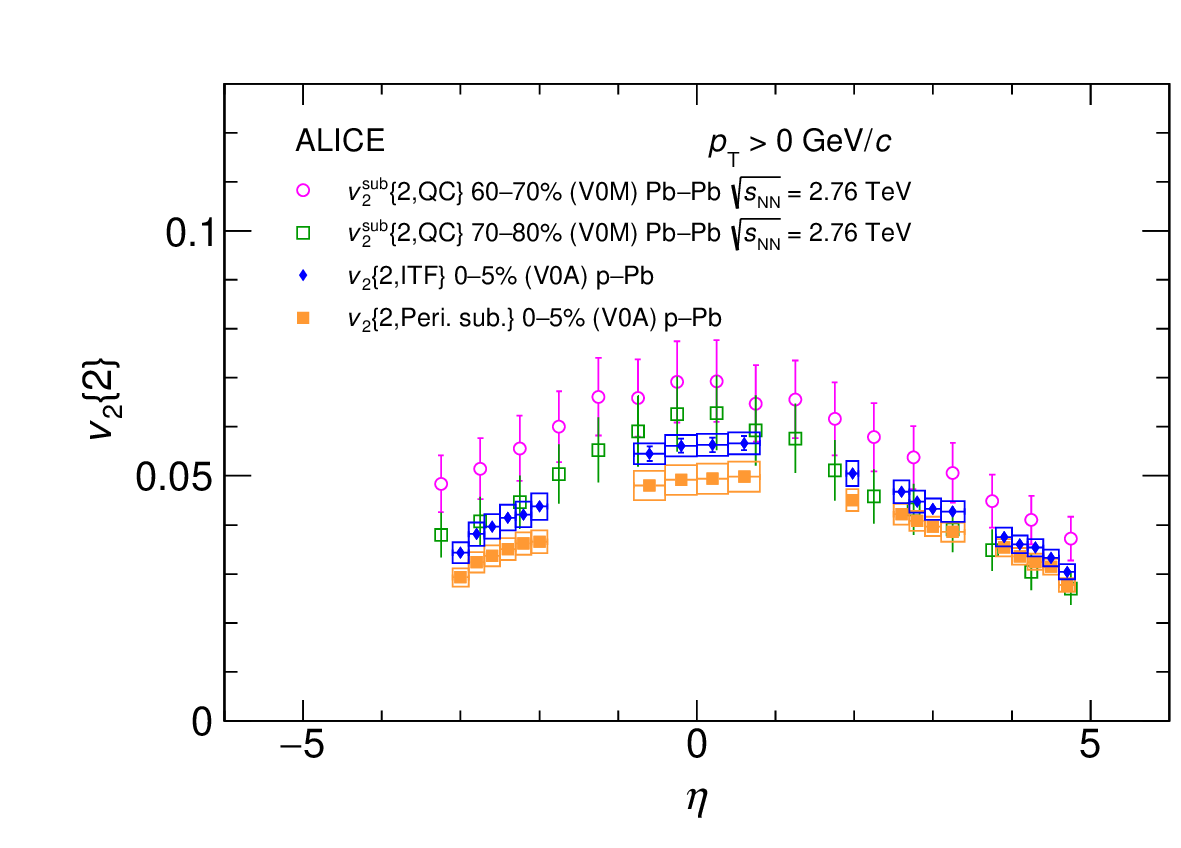}
    \end{center}
    \caption{
    Pseudorapidity dependence of \pt-integrated \vtwo as measured in peripheral Pb--Pb collisions and central p--Pb collisions.
    The Pb–Pb results were obtained with the Q-cumulant method~\cite{ALICE:2016tlx}. The p--Pb results were obtained with the improved-template-fit and peripheral subtraction methods.
    }
    \label{fig:compPbPb2}
\end{figure}

\begin{figure}[H]
    \begin{center}
    \includegraphics[width=0.75\textwidth]{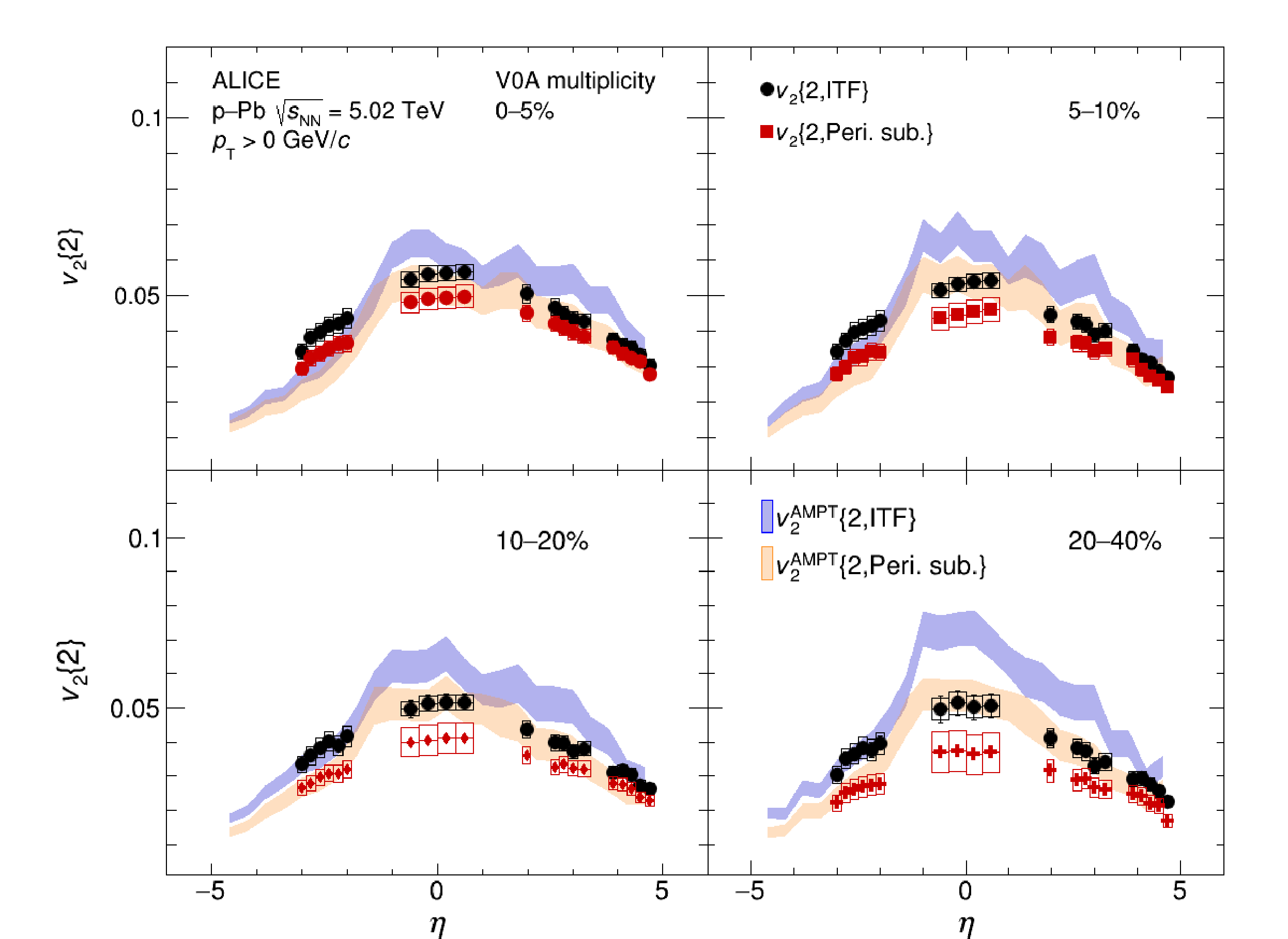}
    \end{center}
    \caption{
    Pseudorapidity dependence of \pt-integrated \vtwo as measured in different p--Pb centrality classes and as obtained from the 
AMPT calculation with the string melting configuration. The \vtwo were extracted using the improved-template-fit method and the peripheral subtraction.}
    \label{fig:amptimp1}
\end{figure}

Figure~\ref{fig:compnonflow} shows the \pt-integrated \vtwo as a function of $\eta$ with two different non-flow suppression methods for 0--5\%, 5--10\%, 10--20\%, and 20--40\% \pPb collisions.
The improved-template-fit method, which considers \vtwo in peripheral collisions, gives smaller \vtwo than the template-fit method; 
however, the difference is insignificant for the presented pseudorapidity regions. The effect of the second-order component in the peripheral collision is small. The result of \vtwo from the peripheral subtraction method, which assumes no elliptic flow in the peripheral collision, is about 15$\%$ smaller than the template procedure for the 0--5$\%$ centrality class. Regardless of which non-flow subtraction method is applied, non-zero \vtwo results are observed for the entire presented pseudorapidity region with more than 5-$\sigma$ confidence, further confirming the observations of anisotropic flow in high-multiplicity \pPb collisions.
Figure~\ref{fig:dndetaappendix} shows the \vtwo as a function of local charged-particle density for five different pseudorapidity regions with the peripheral subtraction (left) and the improved-template-fit (right) methods. Similar to the template fit results shown in Fig.~\ref{fig:dndeta}, \vtwo from peripheral subtraction and improved-templated-fit methods show a charged multiplicity density dependence.
Fig.~\ref{fig:compPbPb2} also shows the comparisons of \vtwo in 
0--5\% central \pPb collisions and in peripheral \PbPb collisions with the improved-template-fit method and the peripheral subtraction. The \vtwo of the improved-template-fit method is comparable to the \vtwo in peripheral \PbPb collisions with similar multiplicity at forward rapidity as well as the \vtwo of the template fit. At the same time, the \vtwo result from the peripheral subtraction is consistent with the results in \PbPb in the Pb-going direction within the uncertainty; however, it is smaller in the p-going direction.
Figure~\ref{fig:amptimp1} compares the AMPT calculation and the results with the improved-template-fit and the peripheral subtraction methods.
Similar to the experimental measurements, improved-template-fit and peripheral subtraction methods are also applied in these AMPT calculations.
It is found that the AMPT calculations also exhibit the differences in \vtwo from the improved-template-fit method and the peripheral subtraction, just as the data. This is because both template-fit and improved-template-fit methods consider possible flow generated in peripheral collisions, which is the case in the AMPT model, while such flow contributions are treated as non-flow and subtracted in the peripheral subtraction method. 

\newpage
\section{The ALICE Collaboration}
\label{app:collab}
\begin{flushleft} 
\small

S.~Acharya\,\orcidlink{0000-0002-9213-5329}\,$^{\rm 126}$, 
D.~Adamov\'{a}\,\orcidlink{0000-0002-0504-7428}\,$^{\rm 86}$, 
G.~Aglieri Rinella\,\orcidlink{0000-0002-9611-3696}\,$^{\rm 33}$, 
M.~Agnello\,\orcidlink{0000-0002-0760-5075}\,$^{\rm 30}$, 
N.~Agrawal\,\orcidlink{0000-0003-0348-9836}\,$^{\rm 51}$, 
Z.~Ahammed\,\orcidlink{0000-0001-5241-7412}\,$^{\rm 134}$, 
S.~Ahmad$^{\rm 16}$, 
S.U.~Ahn\,\orcidlink{0000-0001-8847-489X}\,$^{\rm 71}$, 
I.~Ahuja\,\orcidlink{0000-0002-4417-1392}\,$^{\rm 38}$, 
A.~Akindinov\,\orcidlink{0000-0002-7388-3022}\,$^{\rm 142}$
% ...

\section*{Affiliation Notes}

$^{\rm I}$ Also at: Max-Planck-Institut f\"{u}r Physik, Munich, Germany\\
$^{\rm II}$ Also at: Italian National Agency for New Technologies, Energy and Sustainable Economic Development (ENEA), Bologna, Italy\\
$^{\rm III}$ Also at: Dipartimento DET del Politecnico di Torino, Turin, Italy\\
$^{\rm IV}$ Also at: Department of Applied Physics, Aligarh Muslim University, Aligarh, India\\
$^{\rm V}$ Also at: Institute of Theoretical Physics, University of Wroclaw, Poland\\
$^{\rm VI}$ Also at: An institution covered by a cooperation agreement with CERN\\

\section*{Collaboration Institutes}

$^{1}$ A.I. Alikhanyan National Science Laboratory (Yerevan Physics Institute) Foundation, Yerevan, Armenia\\
$^{2}$ AGH University of Krakow, Cracow, Poland\\
$^{3}$ Bogolyubov Institute for Theoretical Physics, National Academy of Sciences of Ukraine, Kiev, Ukraine\\
$^{4}$ Bose Institute, Department of Physics  and Centre for Astroparticle Physics and Space Science (CAPSS), Kolkata, India\\
$^{5}$ California Polytechnic State University, San Luis Obispo, California, United States\\
$^{6}$ Central China Normal University, Wuhan, China\\
$^{7}$ Centro de Aplicaciones Tecnol\'{o}gicas y Desarrollo Nuclear (CEADEN), Havana, Cuba\\
$^{8}$ Centro de Investigaci\'{o}n y de Estudios Avanzados (CINVESTAV), Mexico City and M\'{e}rida, Mexico\\
$^{9}$ Chicago State University, Chicago, Illinois, United States\\
$^{10}$ China Institute of Atomic Energy, Beijing, China\\
$^{11}$ China University of Geosciences, Wuhan, China\\
$^{12}$ Chungbuk National University, Cheongju, Republic of Korea\\
$^{13}$ Comenius University Bratislava, Faculty of Mathematics, Physics and Informatics, Bratislava, Slovak Republic\\
$^{14}$ COMSATS University Islamabad, Islamabad, Pakistan\\
$^{15}$ Creighton University, Omaha, Nebraska, United States\\
$^{16}$ Department of Physics, Aligarh Muslim University, Aligarh, India\\
$^{17}$ Department of Physics, Pusan National University, Pusan, Republic of Korea\\
$^{18}$ Department of Physics, Sejong University, Seoul, Republic of Korea\\
$^{19}$ Department of Physics, University of California, Berkeley, California, United States\\
$^{20}$ Department of Physics, University of Oslo, Oslo, Norway\\
$^{21}$ Department of Physics and Technology, University of Bergen, Bergen, Norway\\
$^{22}$ Dipartimento di Fisica, Universit\`{a} di Pavia, Pavia, Italy\\
$^{23}$ Dipartimento di Fisica dell'Universit\`{a} and Sezione INFN, Cagliari, Italy\\
$^{24}$ Dipartimento di Fisica dell'Universit\`{a} and Sezione INFN, Trieste, Italy\\
$^{25}$ Dipartimento di Fisica dell'Universit\`{a} and Sezione INFN, Turin, Italy\\
$^{26}$ Dipartimento di Fisica e Astronomia dell'Universit\`{a} and Sezione INFN, Bologna, Italy\\
$^{27}$ Dipartimento di Fisica e Astronomia dell'Universit\`{a} and Sezione INFN, Catania, Italy\\
$^{28}$ Dipartimento di Fisica e Astronomia dell'Universit\`{a} and Sezione INFN, Padova, Italy\\
$^{29}$ Dipartimento di Fisica `E.R.~Caianiello' dell'Universit\`{a} and Gruppo Collegato INFN, Salerno, Italy\\
$^{30}$ Dipartimento DISAT del Politecnico and Sezione INFN, Turin, Italy\\
$^{31}$ Dipartimento di Scienze MIFT, Universit\`{a} di Messina, Messina, Italy\\
$^{32}$ Dipartimento Interateneo di Fisica `M.~Merlin' and Sezione INFN, Bari, Italy\\
$^{33}$ European Organization for Nuclear Research (CERN), Geneva, Switzerland\\
$^{34}$ Faculty of Electrical Engineering, Mechanical Engineering and Naval Architecture, University of Split, Split, Croatia\\
$^{35}$ Faculty of Engineering and Science, Western Norway University of Applied Sciences, Bergen, Norway\\
$^{36}$ Faculty of Nuclear Sciences and Physical Engineering, Czech Technical University in Prague, Prague, Czech Republic\\
$^{37}$ Faculty of Physics, Sofia University, Sofia, Bulgaria\\
$^{38}$ Faculty of Science, P.J.~\v{S}af\'{a}rik University, Ko\v{s}ice, Slovak Republic\\
$^{39}$ Frankfurt Institute for Advanced Studies, Johann Wolfgang Goethe-Universit\"{a}t Frankfurt, Frankfurt, Germany\\
$^{40}$ Fudan University, Shanghai, China\\
$^{41}$ Gangneung-Wonju National University, Gangneung, Republic of Korea\\
$^{42}$ Gauhati University, Department of Physics, Guwahati, India\\
$^{43}$ Helmholtz-Institut f\"{u}r Strahlen- und Kernphysik, Rheinische Friedrich-Wilhelms-Universit\"{a}t Bonn, Bonn, Germany\\
$^{44}$ Helsinki Institute of Physics (HIP), Helsinki, Finland\\
$^{45}$ High Energy Physics Group,  Universidad Aut\'{o}noma de Puebla, Puebla, Mexico\\
$^{46}$ Horia Hulubei National Institute of Physics and Nuclear Engineering, Bucharest, Romania\\
$^{47}$ Indian Institute of Technology Bombay (IIT), Mumbai, India\\
$^{48}$ Indian Institute of Technology Indore, Indore, India\\
$^{49}$ INFN, Laboratori Nazionali di Frascati, Frascati, Italy\\
$^{50}$ INFN, Sezione di Bari, Bari, Italy\\
$^{51}$ INFN, Sezione di Bologna, Bologna, Italy\\
$^{52}$ INFN, Sezione di Cagliari, Cagliari, Italy\\
$^{53}$ INFN, Sezione di Catania, Catania, Italy\\
$^{54}$ INFN, Sezione di Padova, Padova, Italy\\
$^{55}$ INFN, Sezione di Pavia, Pavia, Italy\\
$^{56}$ INFN, Sezione di Torino, Turin, Italy\\
$^{57}$ INFN, Sezione di Trieste, Trieste, Italy\\
$^{58}$ Inha University, Incheon, Republic of Korea\\
$^{59}$ Institute for Gravitational and Subatomic Physics (GRASP), Utrecht University/Nikhef, Utrecht, Netherlands\\
$^{60}$ Institute of Experimental Physics, Slovak Academy of Sciences, Ko\v{s}ice, Slovak Republic\\
$^{61}$ Institute of Physics, Homi Bhabha National Institute, Bhubaneswar, India\\
$^{62}$ Institute of Physics of the Czech Academy of Sciences, Prague, Czech Republic\\
$^{63}$ Institute of Space Science (ISS), Bucharest, Romania\\
$^{64}$ Institut f\"{u}r Kernphysik, Johann Wolfgang Goethe-Universit\"{a}t Frankfurt, Frankfurt, Germany\\
$^{65}$ Instituto de Ciencias Nucleares, Universidad Nacional Aut\'{o}noma de M\'{e}xico, Mexico City, Mexico\\
$^{66}$ Instituto de F\'{i}sica, Universidade Federal do Rio Grande do Sul (UFRGS), Porto Alegre, Brazil\\
$^{67}$ Instituto de F\'{\i}sica, Universidad Nacional Aut\'{o}noma de M\'{e}xico, Mexico City, Mexico\\
$^{68}$ iThemba LABS, National Research Foundation, Somerset West, South Africa\\
$^{69}$ Jeonbuk National University, Jeonju, Republic of Korea\\
$^{70}$ Johann-Wolfgang-Goethe Universit\"{a}t Frankfurt Institut f\"{u}r Informatik, Fachbereich Informatik und Mathematik, Frankfurt, Germany\\
$^{71}$ Korea Institute of Science and Technology Information, Daejeon, Republic of Korea\\
$^{72}$ KTO Karatay University, Konya, Turkey\\
$^{73}$ Laboratoire de Physique Subatomique et de Cosmologie, Universit\'{e} Grenoble-Alpes, CNRS-IN2P3, Grenoble, France\\
$^{74}$ Lawrence Berkeley National Laboratory, Berkeley, California, United States\\
$^{75}$ Lund University Department of Physics, Division of Particle Physics, Lund, Sweden\\
$^{76}$ Nagasaki Institute of Applied Science, Nagasaki, Japan\\
$^{77}$ Nara Women{'}s University (NWU), Nara, Japan\\
$^{78}$ National and Kapodistrian University of Athens, School of Science, Department of Physics , Athens, Greece\\
$^{79}$ National Centre for Nuclear Research, Warsaw, Poland\\
$^{80}$ National Institute of Science Education and Research, Homi Bhabha National Institute, Jatni, India\\
$^{81}$ National Nuclear Research Center, Baku, Azerbaijan\\
$^{82}$ National Research and Innovation Agency - BRIN, Jakarta, Indonesia\\
$^{83}$ Niels Bohr Institute, University of Copenhagen, Copenhagen, Denmark\\
$^{84}$ Nikhef, National institute for subatomic physics, Amsterdam, Netherlands\\
$^{85}$ Nuclear Physics Group, STFC Daresbury Laboratory, Daresbury, United Kingdom\\
$^{86}$ Nuclear Physics Institute of the Czech Academy of Sciences, Husinec-\v{R}e\v{z}, Czech Republic\\
$^{87}$ Oak Ridge National Laboratory, Oak Ridge, Tennessee, United States\\
$^{88}$ Ohio State University, Columbus, Ohio, United States\\
$^{89}$ Physics department, Faculty of science, University of Zagreb, Zagreb, Croatia\\
$^{90}$ Physics Department, Panjab University, Chandigarh, India\\
$^{91}$ Physics Department, University of Jammu, Jammu, India\\
$^{92}$ Physics Program and International Institute for Sustainability with Knotted Chiral Meta Matter (SKCM2), Hiroshima University, Hiroshima, Japan\\
$^{93}$ Physikalisches Institut, Eberhard-Karls-Universit\"{a}t T\"{u}bingen, T\"{u}bingen, Germany\\
$^{94}$ Physikalisches Institut, Ruprecht-Karls-Universit\"{a}t Heidelberg, Heidelberg, Germany\\
$^{95}$ Physik Department, Technische Universit\"{a}t M\"{u}nchen, Munich, Germany\\
$^{96}$ Politecnico di Bari and Sezione INFN, Bari, Italy\\
$^{97}$ Research Division and ExtreMe Matter Institute EMMI, GSI Helmholtzzentrum f\"ur Schwerionenforschung GmbH, Darmstadt, Germany\\
$^{98}$ Saga University, Saga, Japan\\
$^{99}$ Saha Institute of Nuclear Physics, Homi Bhabha National Institute, Kolkata, India\\
$^{100}$ School of Physics and Astronomy, University of Birmingham, Birmingham, United Kingdom\\
$^{101}$ Secci\'{o}n F\'{\i}sica, Departamento de Ciencias, Pontificia Universidad Cat\'{o}lica del Per\'{u}, Lima, Peru\\
$^{102}$ Stefan Meyer Institut f\"{u}r Subatomare Physik (SMI), Vienna, Austria\\
$^{103}$ SUBATECH, IMT Atlantique, Nantes Universit\'{e}, CNRS-IN2P3, Nantes, France\\
$^{104}$ Sungkyunkwan University, Suwon City, Republic of Korea\\
$^{105}$ Suranaree University of Technology, Nakhon Ratchasima, Thailand\\
$^{106}$ Technical University of Ko\v{s}ice, Ko\v{s}ice, Slovak Republic\\
$^{107}$ The Henryk Niewodniczanski Institute of Nuclear Physics, Polish Academy of Sciences, Cracow, Poland\\
$^{108}$ The University of Texas at Austin, Austin, Texas, United States\\
$^{109}$ Universidad Aut\'{o}noma de Sinaloa, Culiac\'{a}n, Mexico\\
$^{110}$ Universidade de S\~{a}o Paulo (USP), S\~{a}o Paulo, Brazil\\
$^{111}$ Universidade Estadual de Campinas (UNICAMP), Campinas, Brazil\\
$^{112}$ Universidade Federal do ABC, Santo Andre, Brazil\\
$^{113}$ University of Cape Town, Cape Town, South Africa\\
$^{114}$ University of Derby, Derby, United Kingdom\\
$^{115}$ University of Houston, Houston, Texas, United States\\
$^{116}$ University of Jyv\"{a}skyl\"{a}, Jyv\"{a}skyl\"{a}, Finland\\
$^{117}$ University of Kansas, Lawrence, Kansas, United States\\
$^{118}$ University of Liverpool, Liverpool, United Kingdom\\
$^{119}$ University of Science and Technology of China, Hefei, China\\
$^{120}$ University of South-Eastern Norway, Kongsberg, Norway\\
$^{121}$ University of Tennessee, Knoxville, Tennessee, United States\\
$^{122}$ University of the Witwatersrand, Johannesburg, South Africa\\
$^{123}$ University of Tokyo, Tokyo, Japan\\
$^{124}$ University of Tsukuba, Tsukuba, Japan\\
$^{125}$ University Politehnica of Bucharest, Bucharest, Romania\\
$^{126}$ Universit\'{e} Clermont Auvergne, CNRS/IN2P3, LPC, Clermont-Ferrand, France\\
$^{127}$ Universit\'{e} de Lyon, CNRS/IN2P3, Institut de Physique des 2 Infinis de Lyon, Lyon, France\\
$^{128}$ Universit\'{e} de Strasbourg, CNRS, IPHC UMR 7178, F-67000 Strasbourg, France, Strasbourg, France\\
$^{129}$ Universit\'{e} Paris-Saclay, Centre d'Etudes de Saclay (CEA), IRFU, D\'{e}partment de Physique Nucl\'{e}aire (DPhN), Saclay, France\\
$^{130}$ Universit\'{e}  Paris-Saclay, CNRS/IN2P3, IJCLab, Orsay, France\\
$^{131}$ Universit\`{a} degli Studi di Foggia, Foggia, Italy\\
$^{132}$ Universit\`{a} del Piemonte Orientale, Vercelli, Italy\\
$^{133}$ Universit\`{a} di Brescia, Brescia, Italy\\
$^{134}$ Variable Energy Cyclotron Centre, Homi Bhabha National Institute, Kolkata, India\\
$^{135}$ Warsaw University of Technology, Warsaw, Poland\\
$^{136}$ Wayne State University, Detroit, Michigan, United States\\
$^{137}$ Westf\"{a}lische Wilhelms-Universit\"{a}t M\"{u}nster, Institut f\"{u}r Kernphysik, M\"{u}nster, Germany\\
$^{138}$ Wigner Research Centre for Physics, Budapest, Hungary\\
$^{139}$ Yale University, New Haven, Connecticut, United States\\
$^{140}$ Yonsei University, Seoul, Republic of Korea\\
$^{141}$  Zentrum  f\"{u}r Technologie und Transfer (ZTT), Worms, Germany\\
$^{142}$ Affiliated with an institute covered by a cooperation agreement with CERN\\
$^{143}$ Affiliated with an international laboratory covered by a cooperation agreement with CERN.\\

\end{flushleft} 
  
\end{document}